\documentclass[nonacm,sigconf]{acmart}

\usepackage{algorithmic}
\usepackage[ruled,linesnumbered,vlined]{algorithm2e}
\usepackage{setspace}
\usepackage{listings}
\usepackage{multicol}
\usepackage{multirow}
\usepackage{makecell}
\usepackage{amsthm}
\usepackage{enumitem}
\usepackage{tikz}
\usepackage{xcolor}
\usepackage{colortbl}
\usepackage[subtle]{savetrees} %
\usepackage[font=footnotesize,labelfont=bf]{subcaption}

\renewcommand\footnotetextcopyrightpermission[1]{}

\newcommand{\parabf}[1]{\noindent\textbf{#1}}

\newcommand{\CodeIn}[1]{{\small \texttt{#1}}}
\newcommand{\Comment}[1]{}

\newcommand{\tech}{\textsc{FuzzGPT}\xspace} %
\newcommand{\techzs}{\tech-ZS\xspace}
\newcommand{\techfs}{\tech-FS\xspace}
\newcommand{\techft}{\tech-FT\xspace}

\newcommand{\numPtTotalAPI}{1593\xspace}

\newcommand{\numTFTotalAPI}{3316\xspace}

\newcommand{\numPtBuggyCode}{1750\xspace}
\newcommand{\numTFBuggyCode}{633\xspace}

\newcommand{\ptCoverage}{33.72\%\xspace}

\newcommand{\ptCoverageImproveZS}{60.70\%\xspace}
\newcommand{\tfCoverage}{54.37\%\xspace} 
\newcommand{\tfCoverageImprove}{36.03\%\xspace}

\newcommand{\numTotalBugs}{76\xspace}
\newcommand{\numTotalConfirmedBugs}{61\xspace}
\newcommand{\numTotalConfirmUnknownBugs}{49\xspace}
\newcommand{\numTotalCrashBugs}{25\xspace}
\newcommand{\numTotalConsistencyBugs}{24\xspace}
\newcommand{\numTotalADRelatedBugs}{30\xspace}
\newcommand{\numTotalFixedBugs}{6\xspace}
\newcommand{\numTotalHighPrioBugs}{11\xspace}
\newcommand{\numPtHighPrioBugs}{3\xspace}
\newcommand{\numTfVulnerability}{8\xspace}

\newcommand{\numTitanFuzzCanFindConfirmedUnknownBugs}{11\xspace}

\newcommand{\numHistoryBugCanFindConfirmedUnknownBugs}{2\xspace}

\newcommand{\muffin}{Muffin\xspace}
\newcommand{\freefuzz}{FreeFuzz\xspace}

\newcommand{\deeprel}{DeepREL\xspace}

\newcommand{\nablafuzz}{$\nabla$Fuzz\xspace} 
\newcommand{\llmfuzz}{\textsc{TitanFuzz}\xspace}
\newcommand{\titanfuzz}{\textsc{TitanFuzz}\xspace}

\newcommand{\langfuzz}{LangFuzz\xspace}
\newcommand{\javatailor}{JavaTailor\xspace}

\newcommand{\tf}{TensorFlow\xspace}
\newcommand{\pt}{PyTorch\xspace}
\newcommand{\python}{Python\xspace}

\newcommand{\gpt}{GPT\xspace} 
\newcommand{\codex}{Codex\xspace} 
\newcommand{\chatgpt}{ChatGPT\xspace} 
\newcommand{\incoder}{\textsc{InCoder}\xspace}
\newcommand{\codegen}{\textsc{CodeGen}\xspace}

\newcommand{\llm}{LLM\xspace}
\newcommand{\llmfull}{Large Language Model\xspace}
\newcommand{\nlp}{NLP\xspace}
\newcommand{\coft}{CoT\xspace}
\newcommand{\ft}{FT\xspace} 
\newcommand{\fs}{FS\xspace} 
\newcommand{\zs}{ZS\xspace}

\newcommand{\finetune}{fine-tune\xspace} 
\newcommand{\fewshot}{few-shot\xspace} 
\newcommand{\zeroshot}{zero-shot\xspace} 
\newcommand{\Finetune}{Fine-tune\xspace} 
\newcommand{\Fewshot}{Few-shot\xspace} 
\newcommand{\Zeroshot}{Zero-shot\xspace} 
\newcommand{\Incontextlearning}{In-context learning\xspace}
\newcommand{\incontextlearning}{in-context learning\xspace}

\newcommand{\kshot}{K\xspace}
\newcommand{\Kshot}{\(K\)-Shot\xspace}
\newcommand{\knum}{6\xspace}

\newcommand{\ad}{AD\xspace}
\newcommand{\nd}{ND\xspace}

\newcommand{\unusual}{unusual\xspace}

\newcommand{\ordinary}{ordinary\xspace}

\newcommand*\circled[1]{\tikz[baseline=(char.base)]{
            \node[shape=circle,fill,inner sep=1pt] (char) {\textcolor{white}{{\footnotesize #1}}};}}

\newcommand{\distance}{5pt}
\begin{document}

\title{Large Language Models are Edge-Case Fuzzers:\\ Testing Deep Learning Libraries via \tech}

\author{Yinlin Deng}
    \affiliation{\institution{University of Illinois\\ Urbana-Champaign}\country{}}
    \email{yinlind2@illinois.edu}
\author{Chunqiu Steven Xia}
    \affiliation{\institution{University of Illinois\\ Urbana-Champaign}\country{}}
    \email{chunqiu2@illinois.edu}
\author{Chenyuan Yang}
    \affiliation{\institution{University of Illinois\\ Urbana-Champaign}\country{}}
    \email{cy54@illinois.edu}
\author{Shizhuo Dylan Zhang}
    \affiliation{\institution{University of Illinois\\ Urbana-Champaign}\country{}}
    \email{shizhuo2@illinois.edu}
\author{Shujing Yang}
    \affiliation{\institution{University of Illinois\\ Urbana-Champaign}\country{}}
    \email{shujing6@illinois.edu}
\author{Lingming Zhang}
    \affiliation{\institution{University of Illinois\\ Urbana-Champaign}\country{}}
    \email{lingming@illinois.edu}

\begin{abstract}

Bugs in Deep Learning (DL) libraries may affect almost all downstream DL applications, and it is crucial to ensure the quality of such systems. It is challenging to generate valid input programs for fuzzing DL libraries, since the input programs need to satisfy both the syntax/semantics of the supported languages (e.g., Python) and the tensor/operator constraints for constructing valid computational graphs. Recently, the \llmfuzz work demonstrates that modern \llmfull{s} (\llm{s}) can be directly leveraged to implicitly learn all the language and DL computation constraints to generate valid programs for fuzzing DL libraries. However, \llm{s} tend to generate \ordinary programs following similar patterns/tokens with typical programs seen in their massive training corpora (e.g., GitHub), while fuzzing favors \emph{\unusual} inputs that cover edge cases or are unlikely to be manually produced. To fill this gap, this paper proposes \tech, the first technique to prime \llm{s} to synthesize \unusual programs for fuzzing. \tech is built on the well-known hypothesis that historical bug-triggering programs may include rare/valuable code ingredients important for bug finding. Meanwhile, while traditional techniques leveraging such historical information require intensive human efforts to both design dedicated generators and ensure the syntactic/semantic validity of generated programs, \tech demonstrates that this process can be \emph{fully automated} via the intrinsic capabilities of \llm{s} (including fine-tuning and in-context learning), while being generalizable and applicable to challenging domains. 
While \tech can be applied with different \llm{s}, this paper focuses on the powerful GPT-style models: \codex and \codegen.
Moreover, \tech also shows the potential of directly leveraging the instruct-following capability of the recent \chatgpt for effective fuzzing. The experimental study on two popular DL libraries (\pt and \tf) shows that \tech can substantially outperform \titanfuzz, detecting \numTotalBugs bugs, with \numTotalConfirmUnknownBugs already confirmed as previously unknown bugs, including \numTotalHighPrioBugs high-priority bugs or security vulnerabilities.%

\end{abstract}

\maketitle

\section{Introduction}

Deep Learning (DL) has been widely adopted in various application domains, including scientific discovery~\cite{raghu2020survey}, healthcare~\cite{healthcare}, finance~\cite{finance}, and transportation~\cite{transportation}. Such DL applications are commonly constructed using DL libraries (e.g., \pt~\cite{PyTorch} and \tf~\cite{Tensorflow}) where developers utilize library APIs to build, train, and deploy DL models. Similar to any other complicated software systems, DL libraries can also be buggy. Moreover, bugs in DL libraries can cause serious consequences as they can potentially affect almost all downstream DL applications, including safety-critical ones~\cite{aiforcancer,aiforroad}. As a result, it is crucial to ensure the quality of DL libraries.

\begin{figure}
    \centering
    \includegraphics[width=\linewidth]{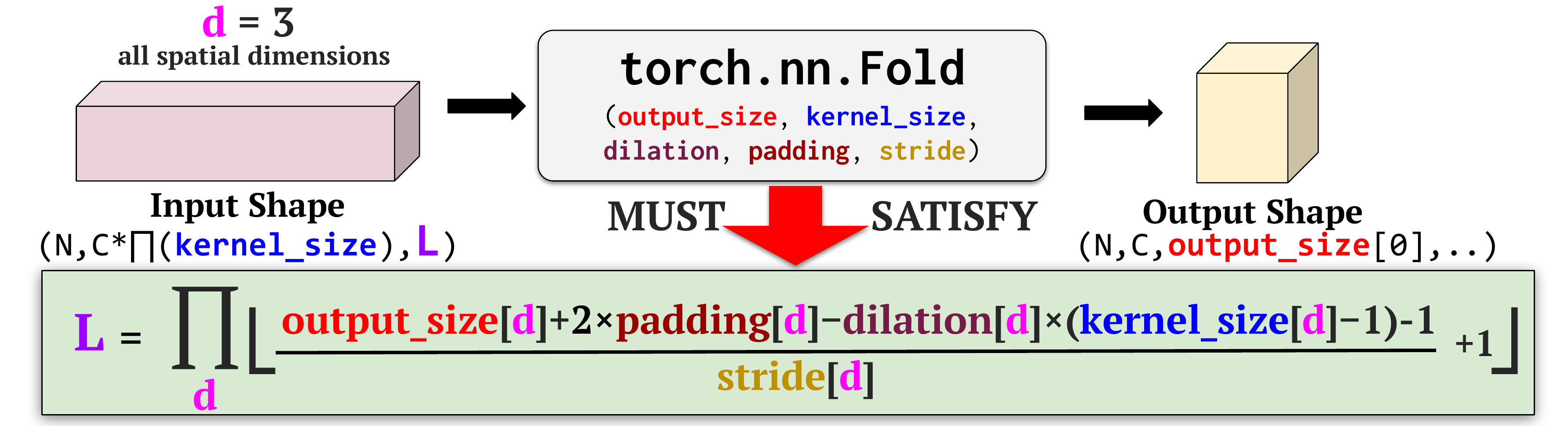}
    \caption{Tensor computation with \CodeIn{torch.nn.Fold}.}
    \label{fig:tensor-comp-fold}
\end{figure}

Fuzzing~\cite{SuttonFuzzingBook, Boehme2021fuzzing, zeller2019fuzzing}, a powerful methodology for bug finding via random input generation, has been widely studied for testing DL libraries in recent years. Meanwhile, it is extremely challenging to generate arbitrary input programs for DL libraries, since the programs need to satisfy both the syntax/semantics of the supported languages (such as Python with dynamic typing) and the tensor/operator constraints for constructing valid computational graphs. For example, Figure~\ref{fig:tensor-comp-fold} shows the complicated constraints for input/output tensor shapes and other operator inputs (e.g., \CodeIn{dilation}) to correctly use a \pt API \CodeIn{torch.nn.Fold}: the last shape value \(L\) in the input should equal the product of the formula across all spatial dimensions \(d\). 
To simplify the problem, prior DL library fuzzing techniques mainly work on model-level~\cite{cradle,audee,lemon,muffin,nnsmith} or API-level fuzzing~\cite{freefuzz,docter,deeprel,nablafuzz}. Model-level fuzzers either re-use/mutate existing seed models~\cite{cradle, audee, lemon}, or generate DL models from scratch~\cite{nnsmith, muffin}. Due to the intricate tensor/operator constraints, such model-level fuzzers either only focus on manipulating shape-preserving APIs~\cite{lemon} or require manually-written specifications for each API~\cite{muffin,nnsmith} to preserve model validity. As a result, they can only cover limited DL APIs and program patterns. On the other hand, API-level fuzzers focus on testing each individual API via effective input generation~\cite{freefuzz, docter} or oracle inference~\cite{deeprel, nablafuzz}. While API-level fuzzers can easily cover a large number of APIs, they cannot find any bugs arising from interactions of different DL APIs. 

With the recent enormous advances in \llmfull{s} (\llm{s}), \titanfuzz~\cite{titanfuzz} has been proposed to directly leverage \llm{s} for fuzzing DL libraries. The key insight is that \llm{s} are pre-trained on billions of code snippets in different languages from open source, which can include numerous valid DL programs for popular DL libraries; in this way, \llm{s} can implicitly learn both the language syntax/semantics and the tensor/operator constraints for valid DL computations. \llmfuzz has been shown effective in generating \emph{valid} input DL programs and substantially outperforms both traditional model-level and API-level fuzzers. More importantly, compared with traditional fuzzing techniques that require intensive human efforts for building generation/mutation strategies~\cite{yang2011csmith, holler2012fuzzing,donaldson2017automated,le2014compiler,le2015finding,sun2016finding,zhang2017skeletal}, \llmfuzz is fully automated, and can be easily generalized to different application domains and programming languages. Meanwhile, \llmfuzz directly leverages the generative capability of \llm{s}, which is based on \emph{token naturalness}~\cite{hindle2012natural} and aims to resemble what they saw in the training corpora. In this way, \llmfuzz can easily generate \emph{human-like} DL programs, which definitely can be valid programs for fuzzing. 
However, given DL libraries have been used by developers all over the world, such \ordinary input programs can hardly help cover additional DL library behaviors/paths, leading to limited fuzzing effectiveness.

\parabf{Our Work.} This paper proposes \tech, the first technique to guide \llm{s} to synthesize \emph{\unusual} input programs for effective fuzzing. \tech is built on the well-known hypothesis that historical bug-triggering programs may include edge-case/valuable code ingredients important for bug finding. In the literature, researchers have proposed various techniques to recompose such code ingredients or insert them into new code contexts for exposing new interesting bugs/vulnerabilities~\cite{holler2012fuzzing,chen2019history, soremekun2020inputs,zhao2022javatailor}. However, such techniques require intensive human efforts to both design such dedicated generators and ensure the syntactic/semantic validity of the generated programs. For example, according to prior work~\cite{zhao2022javatailor, holler2012fuzzing}, even resolving the very common undeclared-identifier issue can be non-trivial. Moreover, it is hard to generalize such techniques to different domains, not to mention the challenging DL library fuzzing problem. In contrast, our key insight is that recent advanced \llm{s} offer a natural, generalizable, and fully automated solution for leveraging such historical programs -- they can be easily \emph{prompted}~\cite{radford2019language} or \emph{fine-tuned}~\cite{radford2018improving} to digest such historical programs, and then generate more \unusual programs that resemble the historical ones and effectively exploit their code ingredients. Compared with traditional techniques on leveraging such historical information, \tech can implicitly learn all the generation constraints (including language syntax/semantics, DL computation constraints, and the new \unusual{ness} constraints) implicitly, and is fully automated. Moreover, while this paper focuses on the challenging problem of DL library fuzzing, the \tech idea is directly generalizable to other application domains and programming languages (e.g., fuzzing for various compilers/interpreters, DB systems, SMT solvers, and more).

To implement \tech, we first construct a dataset of bug-triggering code snippets by mining bug reports from open-source repositories of the target DL libraries. Built on this dataset, \tech includes the following strategies. \textbf{(1) \Incontextlearning}~\cite{liu2023pre}: we provide \llm{s} with either a few examples of historical bug-triggering programs (\fewshot learning~\cite{radford2019language,brown2020gpt3}) or a partial bug-triggering program (\zeroshot learning~\cite{petroni2019language,shin2020autoprompt}) to either generate a new code snippet or to autocomplete the partial code. \textbf{(2) Fine-tuning}~\cite{radford2018improving}: we modify the model weights by training on the extracted historical bug-triggering programs to obtain fine-tuned \llm{s} that are specially designed to generate similar bug-triggering code snippets. From both learning strategies, \tech can \emph{prime} the \llm{s} to generate bug-triggering programs by capturing code ingredients within either the local context examples or fine-tuning dataset.%

To summarize, this paper makes the following contributions:

\begin{itemize}[noitemsep, leftmargin=*, topsep=0pt]
    \item \textbf{Dimension} This paper opens up a new direction for generating \emph{\unusual} input programs for effective fuzzing via \llm{s}. This paper is the first to show that \llm{s} can be easily prompted/fine-tuned to resemble historical bug-triggering programs or even directly follow human instructs to generate \unusual programs for fuzzing real-world systems. Compared with traditional fuzzers for \unusual program generation, \tech is fully automated, generalizable, and applicable to challenging application domains.  %
    \item \textbf{Technique} While our idea is generalizable, we have implemented \tech as an \llm-based fuzzer for DL libraries in this paper. We implement three variants of \tech based on \incontextlearning and fine-tuning: 1) \fewshot learning: a few examples of previous bug-triggering code snippets are provided, 2) \zeroshot learning: a partially complete bug-triggering program is given, and 3) fine-tuning: training a specialized \llm via learning bug-ingredients from the historical programs. While \tech can be applied with different \llm{s}, we build our strategies based on state-of-the-art proprietary and open-source GPT-style \llm{s} for code, \codex~\cite{codex} and \codegen~\cite{codegen}. Moreover, we also build a specific \zeroshot \tech variant by directly leveraging the instruct-following capability of \chatgpt~\cite{chatgpt} without any historical information.
    \item \textbf{Extensive Study} We study all \tech variants on two popular DL libraries (\pt~\cite{PyTorch} and \tf~\cite{Tensorflow}). Our results show that \tech achieves \ptCoverageImproveZS/\tfCoverageImprove higher coverage than state-of-the-art \titanfuzz on \pt/\tf. Overall, \tech found \numTotalBugs bugs on the latest versions of \pt and \tf. \numTotalConfirmUnknownBugs have already been confirmed as new bugs, with \numTotalHighPrioBugs high-priority bugs or security vulnerabilities.
\end{itemize}

\section{Background and Related Work}

\subsection{\llmfull}

\llmfull{s} (\llm{s}) have demonstrated impressive performance across a wide range of \nlp tasks by training on large corpora of text data scraped from the Internet~\cite{gao2020pile}. Recently, researchers have begun adopting \llm{s} for code-related tasks, e.g., via further fine-tuning \llm{s} on code snippets from open-source repositories~\cite{feng2020codebert}. \llm{s} can be classified based on variations of the popular Transformer architecture~\cite{vaswani2017attention} into: \emph{Encoder-only}, \emph{Decoder-only} and \emph{Encoder-Decoder} models. Decoder-only \llm{s} (e.g., \gpt~\cite{brown2020gpt3,openai2023gpt4}, \codex~\cite{codex} and \codegen~\cite{codegen}) focus on autoregressive completion tasks  by learning to predict the probability of the next token given previously generated tokens. Encoder-only (e.g., CodeBERT~\cite{feng2020codebert} and GraphCodeBERT~\cite{guo2021graphcodebert}) and Encoder-Decoder (e.g., CodeT5~\cite{wang2021codet5} and PLBART~\cite{ahmad2021PLBART}) models on the other hand are designed for infilling tasks, where the pre-training objective is to recover masked-out tokens or token spans in the training data by using bi-directional context.

 In order to apply \llm{s} for down-stream applications, there are two main paradigms: \emph{Fine-tuning}~\cite{radford2018improving} and \emph{\Incontextlearning}~\cite{radford2019language,brown2020gpt3}. Fine-tuning is the process of updating the model weights by learning the desired output from the given input of a specific downstream task dataset. The resulting fine-tuned \llm can be treated as specialized models designed to perform a particular task such as code summarization~\cite{ahmad2021PLBART}. Different from fine-tuning, which typically requires large downstream datasets to update the model, \incontextlearning only requires a few examples/demonstration of the downstream task. \Incontextlearning directly uses the pre-trained \llm without any weight modification by constructing an input which contains multiple (i.e., few-shot) examples of input/output demonstrations and then the final task to be solved. Through looking at the context, \llm{s} can directly learn the goal of the specific task and expected output format without fine-tuning. Various techniques have been proposed to improve \incontextlearning, in particular via example selection~\cite{liu-etal-2022-makes,rubin2022learning}, and via multi-step reasoning using Scratchpad~\cite{nye2021show} or Chain-of-Thought~\cite{wei2022chain}. Together, \incontextlearning and fine-tuning are two different approaches to prime \llm{s} for downstream tasks. 

While the approach of \tech is general for different \llm{s}, in this paper we mainly focus on the recent powerful GPT-style models: \codex~\cite{codex} and \codegen~\cite{codegen}, which are state-of-the-art proprietary and open-source models for code, respectively. Moreover, we also directly leverage the instruct-following capability of ChatGPT~\cite{chatgpt} for fuzzing.

\subsection{Fuzzing with Historical Bugs}

Fuzzing with historical bugs has been extensively studied for over a decade. One of the pioneering techniques is \langfuzz~\cite{holler2012fuzzing}: \langfuzz first parses historical tests which have been known to cause incorrect behaviors and extracts individual code fragments; then, \langfuzz will generate code and further replace partial code with extracted code fragments. Researchers have also mined fuzzer configurations (e.g., grammar/statement probabilities) from historical bug-triggering programs for more diverse generation~\cite{chen2019history, soremekun2020inputs}.
More recently, \javatailor~\cite{zhao2022javatailor} leverages extracted code ingredients from historical bugs to synthesize new input programs for Java Virtual Machine (JVM) fuzzing. Besides such generation-based or hybrid approaches, mutation-based fuzzers have also widely utilized such prior bug reports or regression tests as high-quality input seeds for future mutations, targeting C compilers~\cite{le2014compiler}, SMT solvers~\cite{winterer2020unusual}, and database systems~\cite{zhong2020squirrel, wang2021industry}.
Moreover, a recent study~\cite{zhong2022enriching} shows that directly reusing code snippets with minimal modification from historical bug reports can also find interesting bugs in other C/C++ compilers.

\tech directly learns from historical bug-triggering code via fine-tuning and \incontextlearning with \llm{s}. Unlike the prior generation- or mutation-based techniques which require extensive human efforts to build dedicated generators and ensure syntactic/semantic validity of generated programs~\cite{holler2012fuzzing, zhao2022javatailor}, \tech{} is fully automated and generalizable. Moreover, it is extremely challenging to build such traditional generators/mutators for DL programs due to the complicated language and tensor/operator constraints, while \tech can implicitly learn all such constraints using modern \llm{s}.%

\subsection{Fuzzing via Deep Learning}

In addition to traditional fuzzing approaches, researchers have also developed fuzzing tools based on Deep Learning (DL) techniques. SeqFuzzer is network protocol fuzzer built on Long Short-term Memory (LSTM)~\cite{hochreiter1997lstm}, a Recurrent Neural Network (RNN)~\cite{cho-etal-2014-learning}. RNN-based fuzzers have also been used for fuzzing other systems such as OpenCL compilers~\cite{cummins2018compiler}, C compilers~\cite{liu2019deepfuzz}, and PDF file parsers~\cite{learnandfuzz}. Likewise, Montage~\cite{lee2020montage} has been proposed to fuzz JavaScript engines by directly training the tree-based RNN to mutate existing seed ASTs. Moreover, COMFORT~\cite{ye2021automated} proposes to fine-tune the pre-trained GPT-2 model to generate JavaScript programs, and further relies on additional heuristics to generate inputs for the synthesized programs. While the above work leverages the development of DL models for fuzzing, they do not yet study modern \llm{s} for code. 
 Recently, \titanfuzz~\cite{titanfuzz} demonstrates for the first time that modern \llm{s} can be directly leveraged to perform end-to-end fuzzing of real-world systems in a zero-shot manner (without any fine-tuning). \titanfuzz first leverages \codex~\cite{codex} to generate high-quality seed programs and then leverages \incoder~\cite{incoder} with an evolutionary fuzzing strategy to generate diverse code snippets. \titanfuzz has demonstrated state-of-the-art results for fuzzing DL libraries and can find bugs that can only be uncovered with complex API sequences. 

In this work, we propose to leverage the historical bug-triggering code information to further guide \llm{s} towards more effective fuzzing. To our knowledge, this is the first work demonstrating that \llm{s} can easily perform history-driven fuzzing (widely studied for over a decade), while being fully automated, generalizable, and applicable to the challenging domain of DL library fuzzing.%

\section{Approach}

\begin{figure*}[t]
    \captionsetup{justification=centering}
    \centering
    \includegraphics[keepaspectratio=true,width=0.9\textwidth]{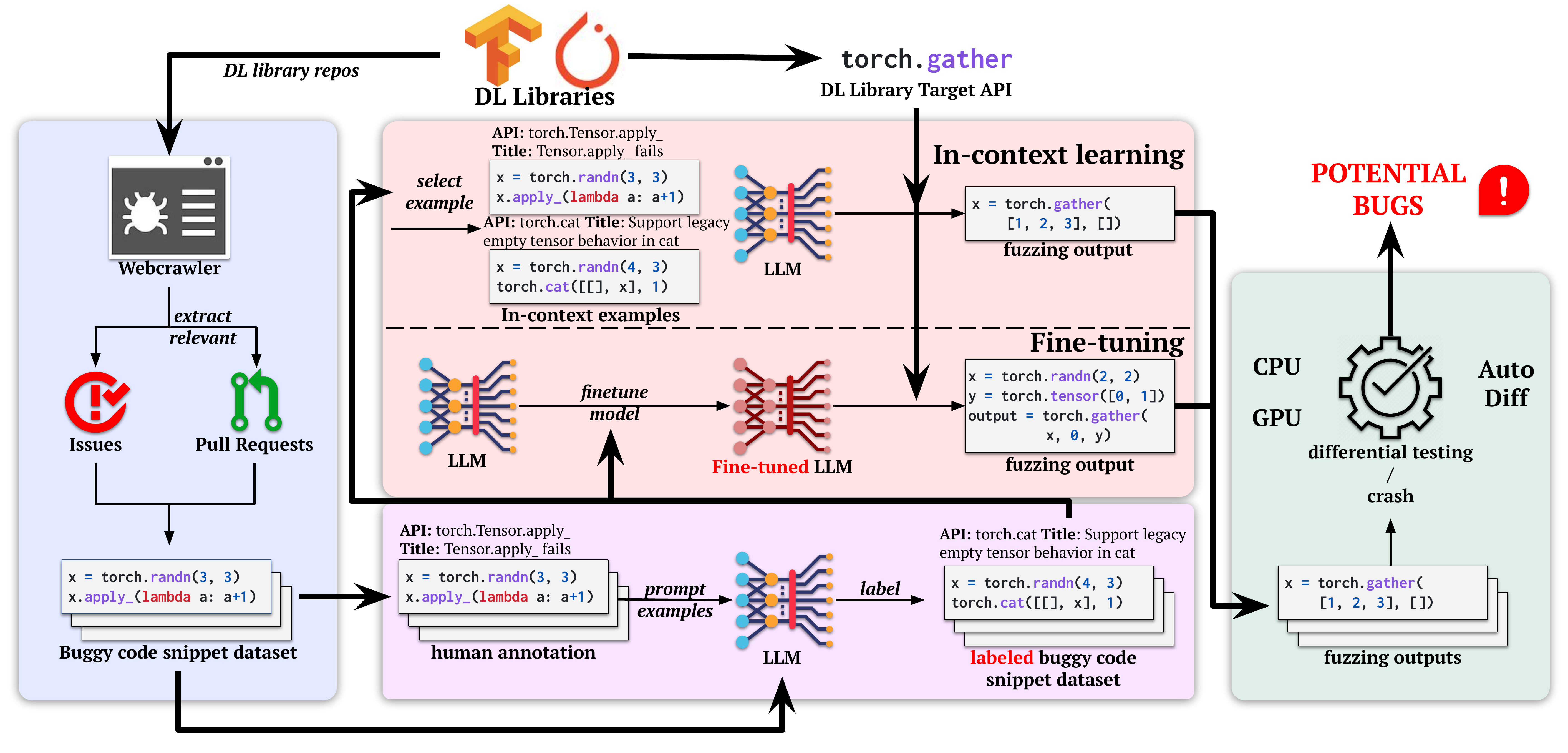}
    \caption{Overview of \tech.}
    \label{fig:overview}
\end{figure*}

In this section, we describe our \tech approach of exploiting historical bugs via \llmfull{s} (\llm{s}) to automatically fuzz DL libraries. The key idea of \tech is to use \llm{s} and directly \emph{learn} from historical reported bugs to generate similar \emph{bug-triggering} code snippets to find new bugs. Existing work along this direction requires extensive human efforts, and can hardly generalize to the challenging domain of DL library fuzzing. In contrast, the recent advances in \llm{s} offer a natural, generalizable, and fully automated solution -- modern \llm{s} can be easily prompted or fine-tuned to digest such historical programs and then generate programs that resemble the historical ones and effectively exploit their code ingredients.  %

Figure~\ref{fig:overview} shows the overview of \tech. We first systematically mine bug reports from the target DL library repositories to collect historical bug-triggering code snippets (Section~\ref{sec:dataset_construction}). As \tech aims to target specific DL library APIs, each bug-triggering code snippet requires a corresponding buggy API label. However, oftentimes the exact buggy API may not be explicitly indicated within the bug report. As such, \tech employs a self-training approach to automatically generate buggy API labels using \llm{s} by prompting with a few manually labeled examples. Next, using these pairs of extracted bug-triggering code snippets and buggy API labels, \tech can start the fuzzing procedure to generate \emph{edge-case} code snippets (Section~\ref{sec:incontext_learning_finetuning}). In this work, we investigate two learning methods: \textbf{1) \Incontextlearning:} we prompt the pre-trained \llm directly with either examples of bug-triggering code and buggy API pairs (\fewshot), or a partial/complete bug-triggering code (\zeroshot), and let the model generate/edit programs for a target API, \textbf{2) Fine-tuning:} using the extracted dataset, we fine-tune the \llm to learn from the bug-triggering code examples and patterns to generate new bug-triggering code snippets for a target API. Finally, the generated programs are executed with test oracles (e.g., CPU/GPU~\cite{freefuzz}, or automatic differentiation oracle~\cite{nablafuzz}) for bug detection (Section~\ref{sec:oracle}).

While \tech is general for any generative \llm{s}, in this paper, we focus on two specific \llm{s}: \codex~\cite{codex} and \codegen~\cite{codegen}. \codex is a state-of-the-art generative model fine-tuned on open-source code repositories initialized from the GPT-3~\cite{brown2020gpt3} model weights. Unlike \codex whose model weights and training data are not released, \codegen is an open-source generative model. Since we target DL library APIs exposed in Python, we use the Python version of \codegen, fine-tuned on Python GitHub code~\cite{codegen}. In \tech, we directly use both \codex and \codegen models, where the larger \codex model allows us to show the full potential of fuzzing when using \llm{s} and the open-source \codegen models can be used to evaluate the scaling effect and test out fine-tuning strategies. We next describe each step in \tech in more detail.

\subsection{Dataset Construction}
\label{sec:dataset_construction}

To facilitate learning from historical bugs, \tech requires a bug-triggering code dataset which contains pairs of bug-exposing code snippet and its corresponding buggy DL library API.%

\subsubsection{Mining Bug History from GitHub.} 
\label{sec:dataset_construct}
First, we implement an HTML crawler to collect all the issues and pull requests (PRs) from the GitHub issue-tracking systems of our target libraries. Then, we focus on identifying bug-triggering code snippets from two sources: (1) Issues associated with accepted or pending PRs. We search in these bug reports for all code blocks, which contain code snippets to reproduce the bug, and concatenate them together. (2) PRs containing code blocks in their commit messages. We further consider PRs because some PRs may fix bugs without corresponding issues. For each extracted issue or PR, we extract its title as well and include that in prompts for \fewshot learning and fine-tuning. After mining bug-triggering code snippets, we will next perform automated annotation to further label each code snippet with a corresponding buggy API.

\subsubsection{Automated Buggy API Annotation}
\label{sec:auto_labeling}

\begin{figure}[h]
    \captionsetup{justification=centering}
    \centering
    \includegraphics[width=\linewidth]{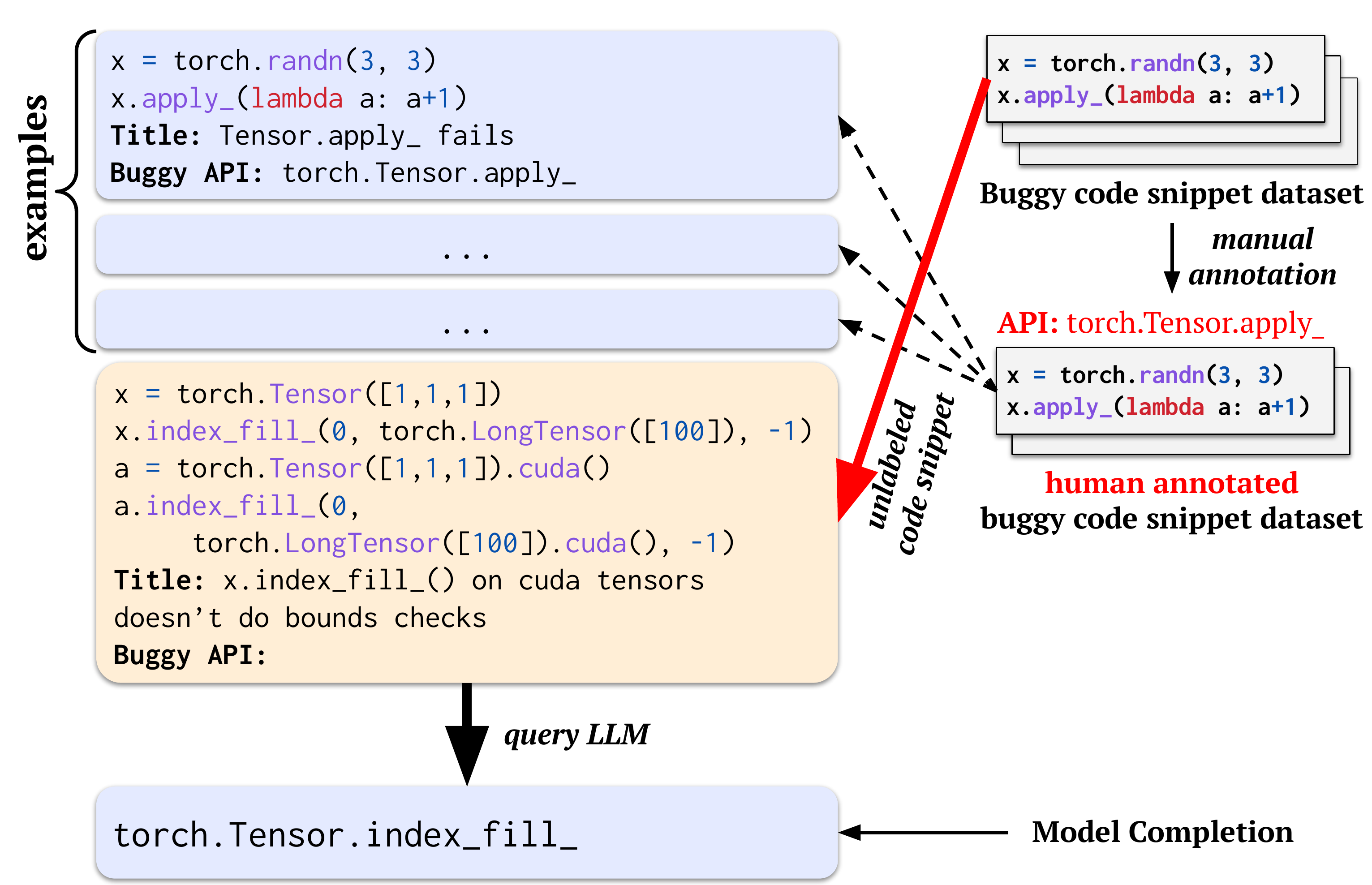}
    \caption{Prompt for buggy API annotation.}
    \label{fig:annotateprompt}
\end{figure}

Each code snippet in our dataset often involves multiple DL APIs. As such, the exact buggy API cannot be directly extracted. In order to annotate the buggy API for each bug-triggering code example, we propose a self-training~\cite{1053799,zoph2020rethinking} approach where we feed a \fewshot prompt (shown in Figure~\ref{fig:annotateprompt}) to \llm and use the model completion as the buggy API. We first provide manual annotation of the buggy API name for several randomly chosen bug-triggering code snippets. These manually annotated pairs become the \fewshot examples used as part of the input to the \llm. In Figure~\ref{fig:annotateprompt}, the \fewshot examples are constructed with the chosen bug-triggering code snippet, the extracted title of the corresponding issue/pull request (e.g., \CodeIn{Tensor.apply\_ fails}), and finally, the manually annotated buggy API name (e.g., \CodeIn{torch.Tensor.apply\_}). Next, we combine the \fewshot examples with a target bug-triggering code snippet to create a prompt, and then query the \llm to obtain the predicted buggy API name for the given code snippet. Note that the annotation effort is \emph{minimal}, since we only need to annotate a few examples (6 for this work as shown in Section~\ref{sec:implementation}) for the targeted library due to the in-context learning capability of modern \llm{s}.

Our proposed method is similar to self-training~\cite{1053799,zoph2020rethinking} where a classifier is first trained on a small labeled dataset and then used to label a larger unlabeled dataset. In our case, we directly use \llm{s} with \fewshot learning from a small number of manual annotations to provide annotations for the large amounts of extracted bug-triggering code snippets in our dataset. Note that the annotation does not need to be fully precise~\cite{min2022rethinking}: the main goal of labeling is to steer the \llm{s} to generate many programs for each targeted DL API by providing paired API and corresponding code examples; that said, even if a code example is unfortunately labeled with a ``wrong'' API which is used in the bug-triggering code for the ground-truth buggy API, it can still guide the model to learn the non-strict mapping from a DL API to a program that contains interesting bug-triggering patterns and hopefully also invokes the API. In our evaluation, we also observe that when targeting a certain DL API, \tech{} can indeed generate code snippets that actually reveal bugs in a different API.

\subsection{\Incontextlearning and Fine-tuning}
\label{sec:incontext_learning_finetuning}

\begin{figure*}[ht]    \captionsetup{justification=centering}
    \centering
    \includegraphics[keepaspectratio=true,width=0.97\textwidth]{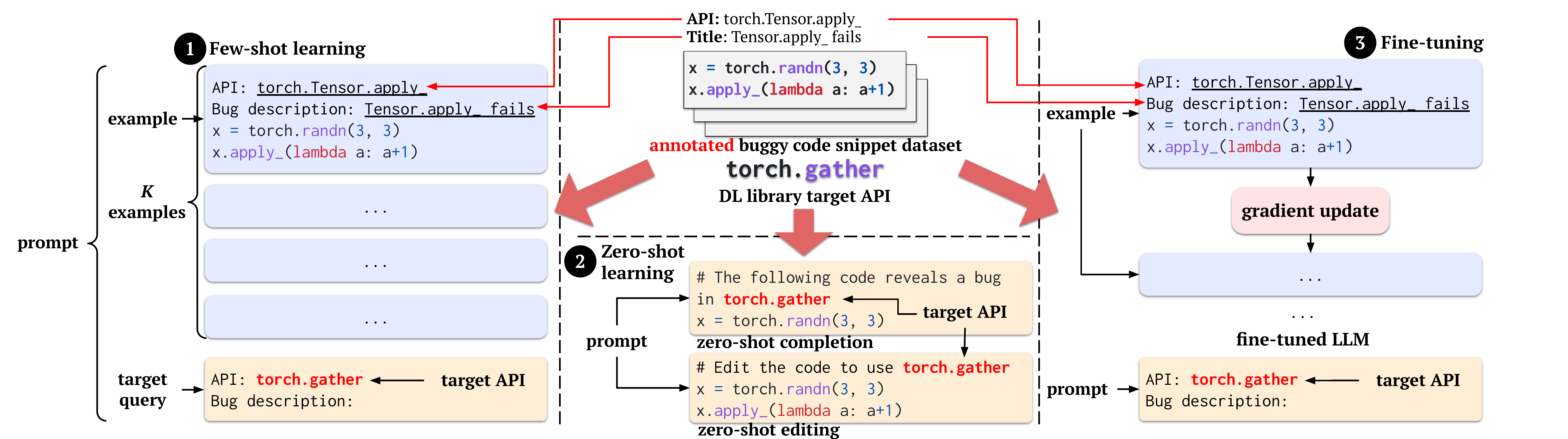}
    \caption{Fine-tuning, \zeroshot, and \fewshot learning for DL Library Fuzzing.}
    \label{fig:prompt}
\end{figure*}

Using the extracted and annotated buggy code snippet dataset, \tech starts the process of \emph{learning} to generate \emph{edge-case} code. At a high level, \llm-based learning methods can be separated into two main approaches: \textbf{\Incontextlearning} and \textbf{Fine-tuning}. \Incontextlearning involves directly using a pre-trained \llm without adjusting any of the model parameters. Instead, the input to the \llm is prepended first with instructions and examples demonstrating the task before providing the current input. \Incontextlearning allows the \llm to prime its output by learning not only the desired output format (e.g., provide a yes or no answer) but also the task domain (e.g., translate English to French) from the surrounding input context. Fine-tuning on the other hand aims to modify the \llm parameters through training on a specific dataset to create a specialized model.

In \tech, we follow prior work~\cite{brown2020language} and systematically explore different learning settings. Specifically, we design three different learning strategies: \fewshot, \zeroshot, and \finetune, each uses the annotated bug-triggering code dataset differently to generate fuzzing outputs. Figure~\ref{fig:prompt} illustrates these three strategies, which are presented in more detail below.%

\subsubsection{\Fewshot Learning}
\label{sec:incontext_learning}

We follow the classic \fewshot \incontextlearning~\cite{brown2020language} by prepending the target query with examples from the annotated bug-triggering code snippet dataset. Figure~\ref{fig:prompt} (\circled{1}) shows how the \fewshot learning prompt is constructed. Our example format consists of the API name (e.g., \CodeIn{torch.Tensor.apply\_}), bug description which is obtained from the title of the issue/PR (e.g., \CodeIn{Tensor.apply\_ fails}), and finally the bug-triggering code snippet. The purpose of these examples is twofold. Firstly, they are intended to prime the \llm towards generating the desired output format. Secondly, they enable the model to learn to produce similar \emph{edge-case} code snippets by observing historical bug-triggering code snippets, without having to modify the model parameters. In \fewshot learning, each prompt consists of \(K\) examples (denoted as \Kshot) and the actual query (the target API and bug description header). Then, \llm{s} can generate new predictions based on the prompt.

\newcommand{\model}{\mathcal{M}\xspace}
\newcommand{\fewshotexamples}{E_{K}\xspace}
\newcommand{\exampledetail}{\{<p_{1}, d_{1}, c_{1}>, ..., <p_{K}, d_{K},c_{K}>\}\xspace}
\newcommand{\exampleprompt}{p_i\xspace}
\newcommand{\exampledescrip}{d_i\xspace}
\newcommand{\examplecode}{c_i\xspace}
\newcommand{\exampleoutput}{o_i\xspace}
\newcommand{\fewshotoutputdescription}{d_{\text{fs}}\xspace}
\newcommand{\fewshotoutputcode}{c_{\text{fs}\xspace}}
\newcommand{\targetprompt}{p_{\text{target}}}

\parabf{Chain-of-Thought Prompting.} An important component of \fewshot learning is the prompt used. Our prompt design is inspired by chain-of-thought (\coft)~\cite{wei2022chain} prompting, where instead of directly generating the final output, the prompt asks the model to perform the task in a step-by-step manner. Following the format of in-context examples, we first include the target API name (e.g., \CodeIn{API: torch.gather}) in our query, and then we ask the model to produce a description of a possible ``bug'' (\CodeIn{Bug description:}) before generating the actual ``bug-triggering'' code snippet that invokes the target API. The predicted bug description provides additional hint to the \llm, indicating that the generated code should try to cover specific potential buggy behavior.

Let $\model$ be the \llm that outputs the probability of generating a sequence. Let $\fewshotexamples = \exampledetail$ be the concatenation of $K$ examples consisting of tuples of example API $\exampleprompt$, bug description $\exampledescrip$, and code snippet $\examplecode$. Let $\targetprompt$ be the target API query. The probability of generating the output code $\fewshotoutputcode$ using \fewshot learning can be formalized as this conditional probability: $\model(\fewshotoutputcode |\fewshotexamples,~  \targetprompt)=\model(\fewshotoutputcode |\fewshotexamples,~  \targetprompt,~ \fewshotoutputdescription)\cdot\model(\fewshotoutputdescription |\fewshotexamples,~  \targetprompt)$

\subsubsection{\Zeroshot Learning}
\label{sec:zeroshot}

We now describe two \tech variants under the \zeroshot learning scenario:

\parabf{\Zeroshot Completion} (default). In this variant, the input is only made up of a partial code snippet. The partial code snippet is created from historical bug-triggering code snippet dataset where we randomly remove a portion of the suffix code. 
Figure~\ref{fig:prompt} (\circled{2}) shows how the \zeroshot completion input can be created. We first include a natural language comment \CodeIn{\# The following code reveals a bug in \{target\_api\}} which allows us to apply \zeroshot learning to any arbitrary target API. We then randomly pick an example bug-exposing code snippet from the dataset. Following, we randomly remove a portion of the selected bug-triggering code snippet's suffix and only keep the prefix lines as the input for the \llm to complete it.

\newcommand{\partialcode}{c_{e}\xspace}
\newcommand{\partialcodej}{j\xspace}
\newcommand{\completionprompt}{p_{\text{comp}}\xspace}
\newcommand{\zeroshotcompletionoutput}{c_{\text{zs-comp}}\xspace}

Again let $\model$ be the \llm that outputs the probability of generating a sequence, $\partialcode$ be the selected code snippet with the first $\partialcodej$ lines ($\partialcode[:\partialcodej]$), and $\completionprompt$ be the completion prompt. The probability of producing \zeroshot completion code $\zeroshotcompletionoutput$ can be formalized as this conditional probability: $\model(\zeroshotcompletionoutput | \completionprompt, \partialcode[:\partialcodej])$, with $\partialcode[:\partialcodej]+\zeroshotcompletionoutput$ being the complete fuzzing code snippet, including the partial code.

\newcommand{\editingprompt}{p_{\text{edit}}\xspace}
\newcommand{\zeroshoteditingoutput}{c_{\text{zs-edit}}\xspace}

\parabf{\Zeroshot Editing.} In this variant, the input is created from a complete code snippet in our historical bug-triggering dataset. Figure~\ref{fig:prompt} (\circled{2}) also shows an example of the \zeroshot editing. To begin with, we use the natural language comment \CodeIn{\# Edit the code to use \{target\_api\}} which indicates to the \llm that we want to perform editing and we should directly reuse a large part of the original code. Similar to \zeroshot completion, we randomly select a bug-triggering code and attach it to the end of the input. Different from \zeroshot completion, where the model autocompletes the end of the code snippet, editing is designed to allow the \llm to reuse all parts of historical bug-triggering code snippet to generate new fuzzing output. Let $\editingprompt$ be the editing prompt. The \zeroshot editing output $\zeroshoteditingoutput$ can be formalized as this conditional probability: $\model(\zeroshoteditingoutput | \editingprompt, \partialcode)$

Compared with \fewshot learning, where multiple historical bugs are provided to the \llm as examples, \zeroshot learning can directly use the existing code portions from historical bugs and only need to generate a partial program (completion) or replace a small portion of the code to use the new target API (edit). These concrete code lines from prior bug-triggering code can be useful to test new APIs as they can include special values (e.g., \CodeIn{NaN}), edge-case tensor shapes/dimensions, and other code ingredients useful for bug finding (e.g., unconventional API usages). By using \zeroshot learning, \tech can directly make use of these historical bug-triggering code snippets to target additional APIs. 

\subsubsection{Fine-tuning}
\label{sec:fine-tuning}

Apart from \incontextlearning (\fewshot and \zeroshot) to learn from historical bugs which only uses the original pre-trained \llm{s}, fine-tuning directly modifies the model parameters by training on the historical bug code snippet dataset. Figure~\ref{fig:prompt} (\circled{3}) shows how the fine-tuning procedure works and also the input to the fine-tuned model during inference time to produce the fuzzing outputs. Each training sample follows the same format as \fewshot examples -- made up of the API name, bug description and also the bug-triggering code snippet. We start with the original pre-trained \llm and then update the model weights through gradient descent by training to auto-regressively predict each training sample.

Let \(T = \{t_1, t_2, ..., t_n\}\) be the training token sequence obtained by tokenizing the training sample, \(T_{<s} = \{t_1, t_2, ..., t_{s-1}\}\) be the token sequence generated by the model so far and \(\model\) be the \llm which outputs the probability of generated the next token given the previously generated tokens. The fine-tuning loss function is defined as: $ \mathcal{L}_{\text{fine-tune}} = -\frac{1}{n}\sum_{i=1}^{n}log\;(\model\;(t_i\;|\;T_{<i}))$

\newcommand{\modelfinetuned}{\mathcal{M}_{\text{ft}}\xspace}
\newcommand{\finetuneoutputcode}{c_{\text{ft}}\xspace}
\newcommand{\finetuneoutputdescription}{d_{\text{ft}}\xspace}

For each DL library, we fine-tune a separate model using the bug-triggering code snippet collected from that library. By fine-tuning on these historical bug-triggering code snippets, the \llm can learn from different kinds of bug-triggering patterns/ingredients for the targeted library. The input to the fine-tuned model follows the same pattern as the \fewshot approach where we construct a prompt based on the specific target API. Let $\modelfinetuned$ be the fine-tuned \llm which outputs the probability of generating a sequence with target API prompt $\targetprompt$. The \finetune model output code $\finetuneoutputcode$ can be formalized as this conditional probability: $\modelfinetuned(\finetuneoutputcode |\targetprompt)=\modelfinetuned(\finetuneoutputcode |\targetprompt,~ \finetuneoutputdescription)\cdot\modelfinetuned(\finetuneoutputdescription |~\targetprompt)$

\subsection{Oracle}
\label{sec:oracle}

Using the generated fuzzing outputs, we test the DL libraries through both general and DL-specific oracles:%
 
\parabf{Crashes.} We detect bugs caused by unexpected crashes found when executing the fuzzing output. These crashes can include aborts, segmentation faults, and \CodeIn{INTERNAL\_ASSERT\_FAILED}. Bugs exposed by such crashes may even further trigger security vulnerabilities. 

\parabf{CPU/GPU oracle.}  We detect wrong-computation bugs by identifying inconsistencies between the output values across two execution backends (CPU and GPU). We follow prior work~\cite{freefuzz,titanfuzz} to use a significance tolerance threshold for comparison in order to account for the non-deterministic nature of particular library APIs when executed on different backends.%

\parabf{Automatic Differentiation (AD) oracle.} We detect gradient computation bugs in the crucial \ad engines of DL libraries, which support efficient training of DL models. We apply the \ad oracle proposed in prior work ~\cite{nablafuzz} and compare the computed gradient between reverse-mode AD (the most commonly used mode in DL libraries), forward-mode AD, and numerical differentiation (\nd).%

\section{Implementation}
\label{sec:implementation}

\parabf{Dataset construction. } 
We scraped the GitHub repositories of targeted libraries using the \emph{requests} library~\cite{requests}, and finally collected \numPtBuggyCode and \numTFBuggyCode bug-triggering code snippets for \pt and \tf, respectively, from their historical issues or PRs. The lower number of \tf is because it has fewer PRs than \pt, as \tf developers are not as active in confirming bugs/PRs, and rarely include code blocks in their PRs. Note that we perform additional cleaning on the extracted code to filter out error messages and remove code lines that contain only the inputs and outputs of executions. We also did not consider the code snippets that fail to pass syntax checking or are longer than 256 tokens. %

For buggy API annotation, we manually annotate $\kshot=\knum$ randomly sampled examples and use them as in-context examples. For each unlabeled example, we query \codex to complete the buggy API with $temperature=0$ (deterministic greedy decoding) to get the most confident prediction following ~\cite{wang2022code4struct}. We also evaluate the labeling accuracy on a random sample of 100 \pt issues/PRs and find the automated labeling achieves 76\% precision. In fact, even mislabeled examples (where the mislabeled API always appears in the code) can still guide \techfs to substantially outperform the zero-shot baseline where we do not have access to any examples (Section~\ref{sec:ablationfewshot}).

\parabf{Language models.} We evaluate on two specific generative \llm{s} \codex(code-davinci-002) and \codegen(350M/2B/6B-mono).  Unlike \codex whose training data and model weights are not released, \codegen~\cite{codegen} is a widely-used open-source generative model~\cite{ding2022cocomic, xia2023conversational, madaan2023learning}, which provides trained model of various sizes and allows fine-tuning. We access \codex through its API and use the PyTorch implementation of the \codegen models on Hugging Face~\cite{HuggingFaceWebPage}.

\parabf{Fine-tuning.} We perform fine-tuning on the \codegen models because \codex is not open source. We fine-tune a separate model for each studied DL library. To update the model parameters for each targeted library, we use batch\_size=32, learning rate=5e-5, and train the models with AdamW optimizer~\cite{loshchilov2018decoupled} for 10 epochs, using a linear learning rate scheduler with 10\% warmup proportion.

\section{Evaluation} 

\subsection{Research Questions}
We investigate the following research questions in our experiments:
\begin{itemize}[noitemsep, leftmargin=*, topsep=0pt]
    \item {\textbf{RQ1:} How does different learning paradigms of \tech compare against each other?}
    \item {\textbf{RQ2:} How does \tech compared against existing fuzzers?}
    \item {\textbf{RQ3:} How do the key components of \tech contribute to its effectiveness? }
    \item {\textbf{RQ4:} Is \tech able to detect new bugs?}
\end{itemize}

\subsection{Experimental Setup}

\parabf{Subject systems.} We evaluate \tech{} on fuzzing \pt and \tf, two of the most popular open-source DL libraries. For RQ1, we separately run \tech with \fewshot, \zeroshot, and \finetune settings to evaluate their individual effectiveness for fuzzing (denoted as \techfs{}, \techzs{}, and \techft{} respectively). We use \techfs as the default implementation for \tech{} if not specified explicitly. For RQ2, We compare our approach with prior work on \pt (v1.12) and \tf (v2.10), the same version as the most recent work \titanfuzz{}\cite{titanfuzz}, and we use the same set of public \python APIs as \titanfuzz{} as well. For RQ3, we run all the generated tests on the nightly version of \pt and \tf to find previously unknown bugs (Section~\ref{sec:bug}). For RQ4, due to huge costs, we conducted our ablation study experiments for all three settings of \tech{} on $50$ APIs randomly sampled from one example DL library \pt, and report the average results of 5 runs following prior work~\cite{titanfuzz}.%

\parabf{Baselines. } We compare \tech{} to state-of-the-art DL library fuzzers, including state-of-the-art API-level (\freefuzz~\cite{freefuzz}, \deeprel ~\cite{deeprel}, and \nablafuzz ~\cite{nablafuzz}) and model-level (\muffin ~\cite{muffin}) fuzzers, as well as the most recent \titanfuzz~\cite{titanfuzz}. We run each tool with its default configuration on both libraries, except that \muffin was only executed on \tf since it does not support \pt. 

\parabf{Environment. } We use a 64-core workstation with 256 GB RAM and running Ubuntu 20.04.5 LTS with 4 NVIDIA RTX A6000 GPUs.

\parabf{Fuzzing budget.} Our default setting generates 100 programs for each target API. In the \Fewshot{} approach, for each target API, we independently construct 10 prompts with 6-shot examples picked randomly, and feed each prompt to the LLM to sample 10 generations. Similarly, in the \zeroshot{} approach, we randomly choose 10 different examples from our dataset, and use the partial/complete code to construct 10 prompts to perform completion/editing. In the \Finetune{} approach, we use a fixed task description, and query the model for 10 times to generation all 100 programs for a target API. 

\parabf{Generation.} Our default setting when using all LLMs for generation uses top-p sampling with $p=0.95$, $temperature=0.8$, and $max\_token=256$ following prior work ~\cite{titanfuzz}.

\subsection{Metrics}

\parabf{Detected bugs.}  Following prior work on DL library fuzzing ~\cite{cradle,lemon,freefuzz,docter,eagle,nablafuzz,titanfuzz}, we report the number of unique detected bugs.

\parabf{Unique crashes.} Besides counting all bugs, we also count the number of unique crashes as another proxy for fuzzing effectiveness. Unique crashes are widely used in the literature for evaluating fuzzing technique~\cite{afl,klees2018evaluating,aflplusplus}.%

\parabf{Code coverage.} Code coverage has been widely adopted in software testing and recently DL library/compiler testing~\cite{freefuzz, muffin, tzer, nablafuzz, titanfuzz}. We follow recent DL library fuzzing work~\cite{muffin, freefuzz, nablafuzz, titanfuzz} and measure Python line coverage with the \emph{coverage.py} tool~\cite{coverage-py}. We excluded additional coverage added by the oracle checking for fair comparison.

\parabf{API coverage.}  We evaluate the number of covered DL APIs as an important metric of test adequacy following prior work on fuzzing DL libraries~\cite{freefuzz, deeprel, nablafuzz, titanfuzz}. 

\parabf{Unique valid programs.} A generated program is considered valid if the program executes successfully without exceptions and actually invokes the target API at least once. We also remove the programs already generated and only consider unique programs. %

\section{Result Analysis}
\label{sec:result}

\subsection{Comparison of Learning Paradigms.} 
We first compare all our three \tech variants (\fewshot, \zeroshot, \finetune) against each other to understand their performance. Table~\ref{tab:comp-paradigm-full} summarizes the results. Columns \textbf{\#APIs}, \textbf{\#Prog.}, and \textbf{Cov} present the number of APIs, unique programs, and lines covered. More specifically, \textbf{Valid} means only unique programs without runtime errors are considered, \textbf{All} means all generated unique programs are considered. Lastly, \textbf{Valid(\%)} computes the ratio of valid programs over all generated unique programs. Figure~\ref{fig:cov-trend-full} further shows the coverage trend with respect to the number of generated programs per API. 

We can observe that \techfs has the highest number of covered APIs and unique (valid) programs on both \pt and \tf. The reason could be that it provides the \llm (i.e., \codex) with a rich context (including \kshot = 6 bug examples, which very likely contain different buggy APIs), enabling it to learn and combine a variety of bug patterns and use a diverse set of APIs.%

\techzs has a much lower valid rate compared with other variants. The reason is that it is required to complete existing partial programs. In this way, the search space is more constrained compared to other variants, and the task is more challenging since the newly generated code needs to be compatible. Meanwhile, \techzs may trigger more interesting interactions between APIs in the existing partial programs and newly generated APIs. Furthermore, the partial code reused from bug history can also be very valuable, and may already cover interesting program paths/behaviors. As a result, \techzs even achieves the highest coverage on \pt. \techzs performs relatively worse on \tf, potentially because there are fewer snippets (only \numTFBuggyCode, which can hardly cover all \numTFTotalAPI \tf APIs) to reuse for \tf.

\techft achieves comparable code coverage on both libraries, and has the highest valid rate on \pt, even with a smaller model (\codegen-6B). The results suggest that fine-tuning can be a very effective approach for fuzzing a specific library, since the fine-tuned model has learned from all collected buggy patterns via updating model parameters, and can ``select'' or ``mix'' the learned buggy ingredients to target a specific API during generation. 
On the contrary, \fewshot{} consumes a limited number of in-context examples, and \zeroshot{} relies on one partial example at each inference step. Nevertheless, fine-tuning requires collecting a (high-quality) fine-tuning dataset, and training a different \llm for every different task (which can be costly in terms of computation resources and storage).

\tech demonstrates the ability to generate fuzzing inputs using techniques from both \incontextlearning and fine-tuning. From the coverage trend in Figure~\ref{fig:cov-trend-full}, we observe that in all three variants, the coverage does not saturate even after all 100 code snippets get generated, showing the power of \llm{s} in learning from historical bug-triggering datasets to continuously generate valuable fuzzing programs that can obtain more coverage. Next we will examine how \tech compares against state-of-the-art DL library fuzzers.

\begin{table}[!ht]\centering
\caption{Comparison of learning paradigms.}\label{tab:comp-paradigm-full}
\scriptsize
\begin{tabular}{l|c|rr|rr|r|rr}\toprule
\multirow{2}{*}{} &\multirow{2}{*}{\textbf{Paradigm}} &\multicolumn{2}{c|}{\textbf{\# APIs}} &\multicolumn{2}{c|}{\textbf{\# Prog.}} &\multirow{2}{*}{\textbf{Valid(\%)}} &\multirow{2}{*}{\textbf{Cov}} \\\cmidrule{3-6} 
& &\textbf{Valid} &\textbf{All} &\textbf{Valid} &\textbf{All} & & \\\midrule
\multirow{3}{*}{\pt} &\techfs &\textbf{1377} &\textbf{1588} &\textbf{42496} &\textbf{154904} &27.43\% &35426 \\
&\techzs &1237 &1553 &7809 &132111 &5.91\% &\textbf{38284} \\
&\techft &1223 &1546 &31225 &112765 &\textbf{27.69\%} &36463 \\ \midrule
\multirow{3}{*}{\tf} &\techfs &\textbf{2309} &\textbf{3314} &\textbf{54058} &\textbf{310483} &\textbf{17.41\%} &\textbf{146487} \\
&\techzs &1460 &3157 &4650 &233887 &1.99\% &126193 \\
&\techft &1834 &3292 &31105 &253216 &12.28\% &125832 \\
\bottomrule
\end{tabular}
\end{table}

\subsection{Comparison with Prior Work}

We compare \tech{}-\fs/-\zs/-\ft against state-of-the-art fuzzer \titanfuzz{} and other recent DL library fuzzers. 
All techniques are applied under their default configurations.

\parabf{API and code coverage.} As shown in Table ~\ref{tab:comp-prior-cov}, all three variants of \tech{} significantly outperform all existing fuzzers including state-of-the-art \titanfuzz{} in code coverage. In particular, the best-performing variants \techfs/\techzs achieve state-of-the-art results of \tfCoverage/\ptCoverage line coverage on \tf/\pt, \tfCoverageImprove/\ptCoverageImproveZS improvement over \titanfuzz{}. We also observe an interesting fact that \tech has similar API coverage with \titanfuzz{} but has much higher code coverage. This demonstrates that \tech can cover much more interesting code behaviors/paths for DL libraries. Both \tech{} and \titanfuzz{} rely on \llm{s} to fully automatically generate (or mutate) programs and significantly outperform prior techniques (\freefuzz, \deeprel, \nablafuzz, \muffin) in terms of API coverage, showing the superiority of \llm{s} for fuzzing.

\begin{figure}[t]
\begin{subfigure}[b]{0.49\columnwidth}
    \centering
    \includegraphics[width=0.95\textwidth]{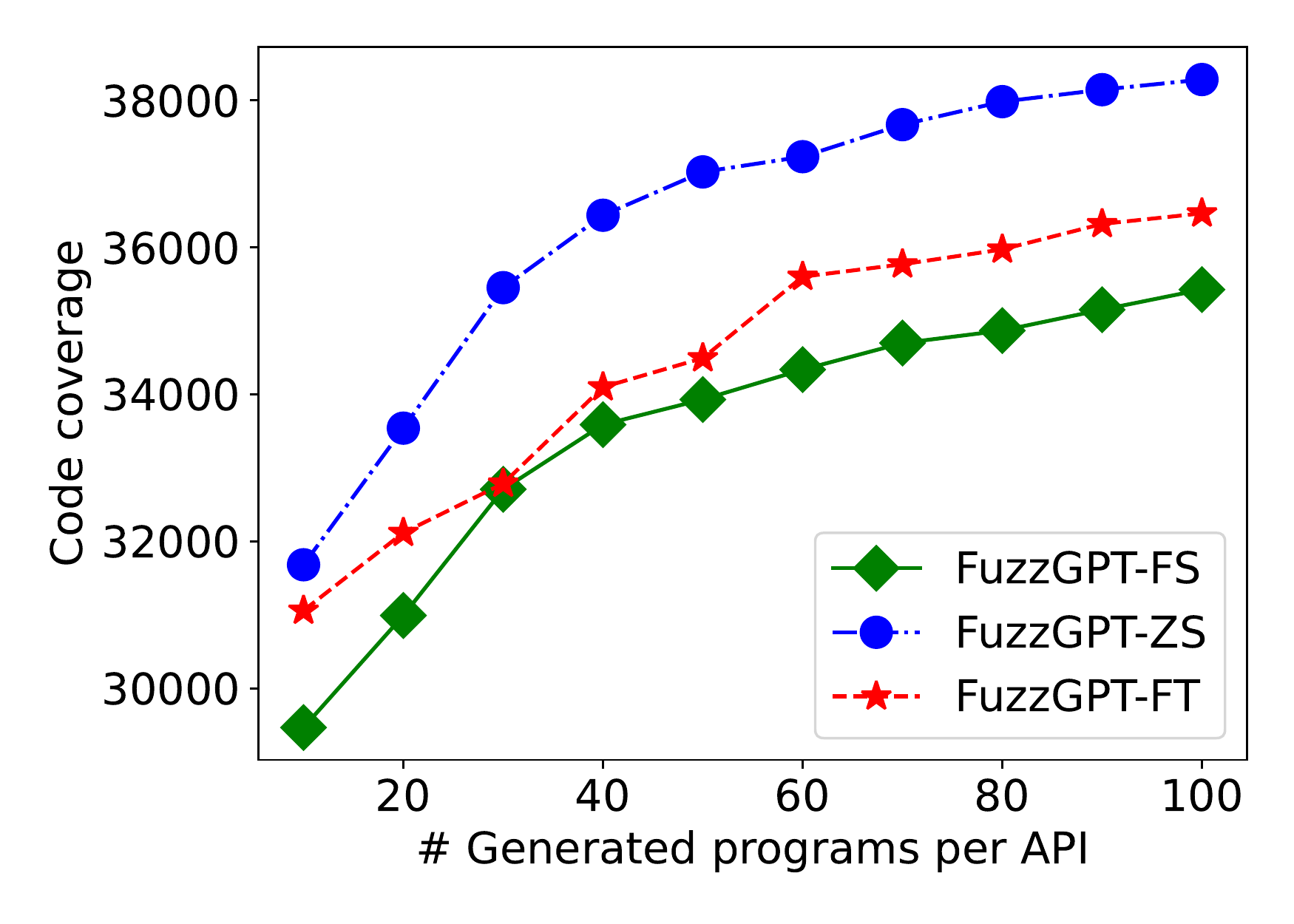}
    \caption{\pt}
    \label{fig:cov-trend-torch}
\end{subfigure}
\begin{subfigure}[b]{0.49\columnwidth}
    \centering
    \includegraphics[width=0.95\textwidth]{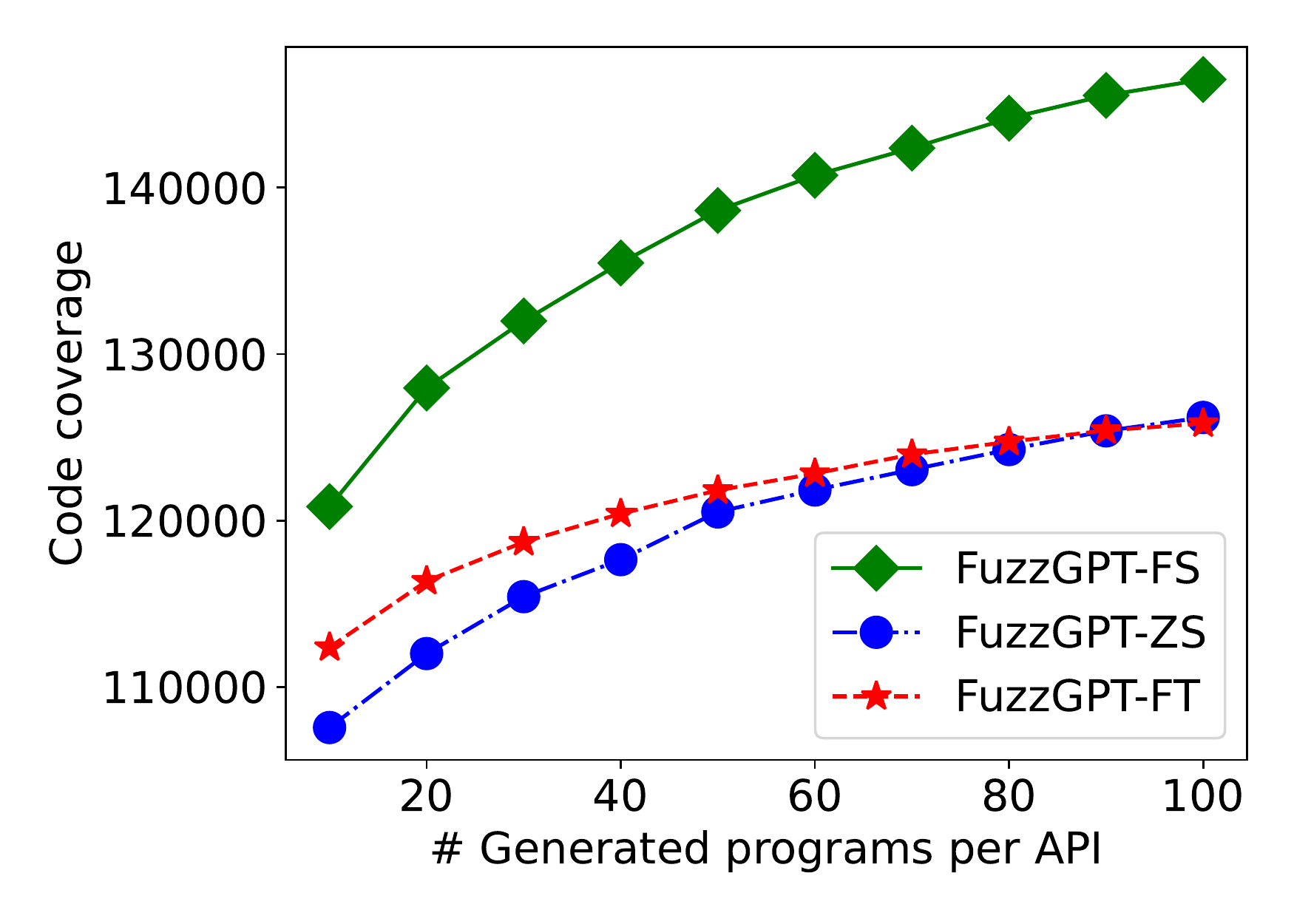}
    \caption{\tf}
    \label{fig:cov-trend-tf}
\end{subfigure}

\caption{Coverage trend of \tech-\fs/-\zs/-\ft.}
\label{fig:cov-trend-full}
\end{figure}

\newcommand{\titanfuzzseed}{\titanfuzz{-seed-only}}

\begin{table}[!htp]\centering
\caption{Comparison with prior work.}\label{tab:comp-prior-cov}
\scriptsize
\begin{tabular}{l|rr|rrr}\toprule
&\multicolumn{2}{c|}{\textbf{\pt}} &\multicolumn{2}{c}{\textbf{\tf}} \\\cmidrule{2-5}
&\textbf{Code Cov} &\textbf{API Cov}&\textbf{Code Cov} &\textbf{API Cov} \\\midrule
Codebase Under Test &113538 (100.00\%) &\numPtTotalAPI &269448 (100.00\%) &\numTFTotalAPI \\ \midrule
\freefuzz &15688 (13.82\%) &468 &78548 (29.15\%) &581 \\
\deeprel &15794 (13.91\%) &1071 &82592 (30.65\%) &1159 \\
\nablafuzz &15860 (13.97\%) & 1071 & 89722 (33.30\%) &1159 \\
\muffin &NA &NA &79283 (29.42\%) &79 \\
\titanfuzzseed &22584 (19.89\%) &1329 &103054 (38.35\%) &2215 \\
\titanfuzz{} &23823 (20.98\%) &1329 &107685 (39.97\%) &2215 \\ \midrule
\tech{}-\fs-25 & 32305 (28.45\%) & 1296 & 130312 (48.36\%) & 1937 \\
\tech{}-\fs &35426 (31.20\%) & {\bf 1377} & {\bf 146487 (\tfCoverage)} & {\bf 2309}\\
\tech{}-\zs & {\bf 38284 (33.72\%)} & 1237 & 126193 (46.83\%) & 1460\\
\tech{}-\ft &36463 (32.12\%) & 1223 & 125832 (46.70\%) & 1834\\ 
\bottomrule
\end{tabular}
\end{table}

\begin{figure}[t]
\begin{subfigure}[b]{0.49\columnwidth}
    \centering
    \includegraphics[width=0.95\textwidth]{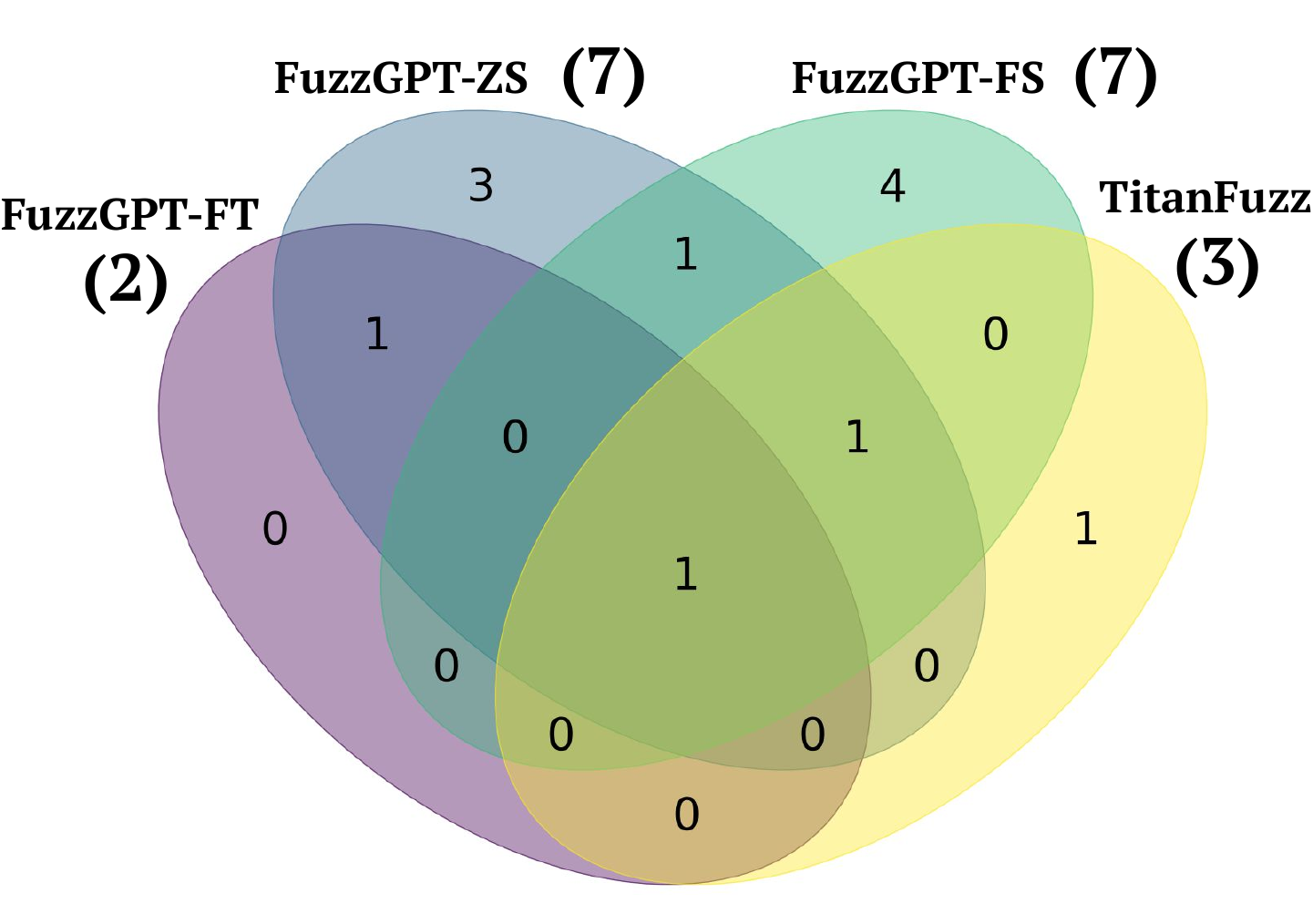}
    \caption{Direct crashes.}
    \label{fig:venn-direct-crash-torch}
\end{subfigure}
\begin{subfigure}[b]{0.49\columnwidth}
    \centering
    \includegraphics[width=0.95\textwidth]{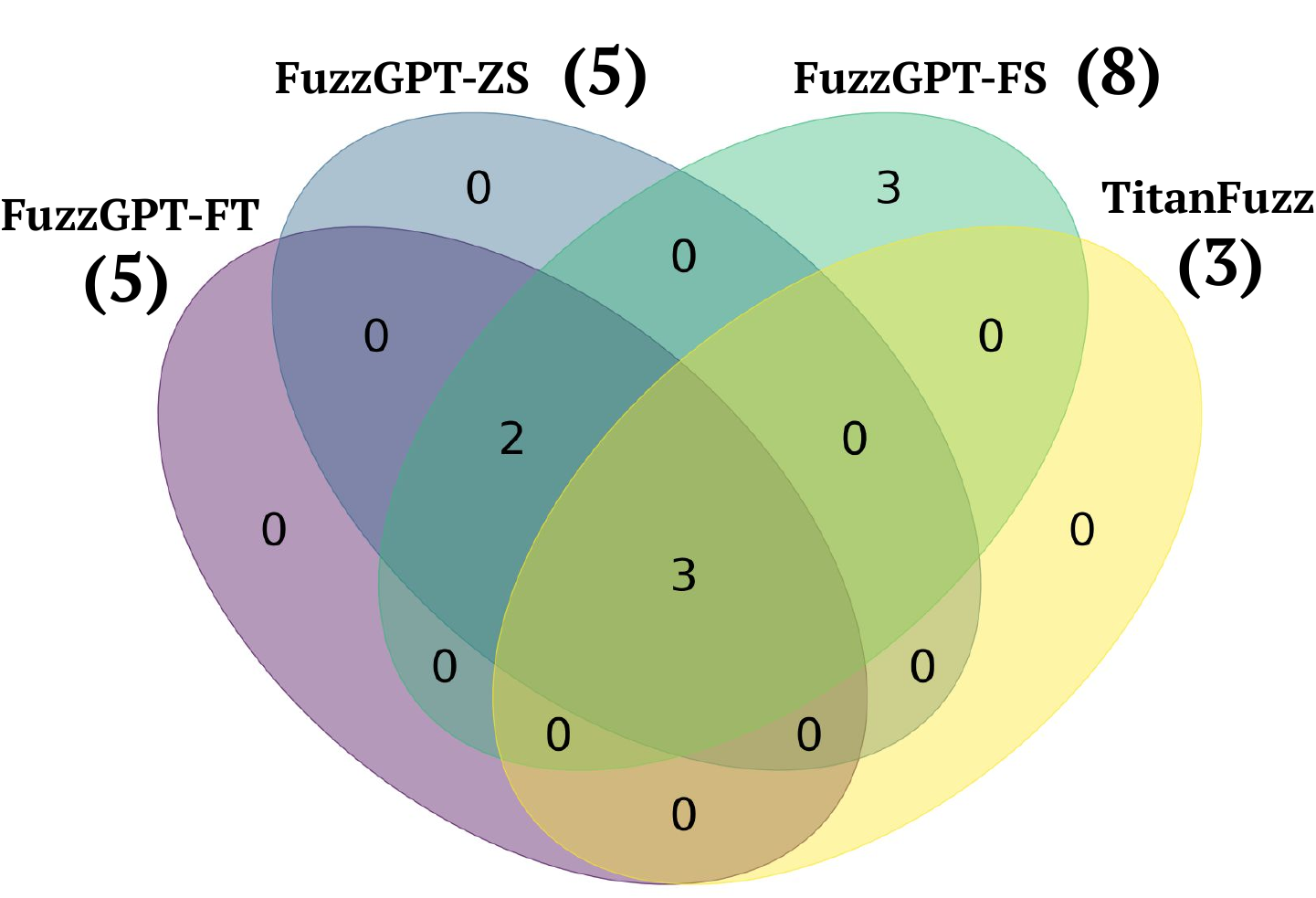}
    \caption{\ad Crashes.}
    \label{fig:venn-ad-crash-torch}
\end{subfigure}

\caption{Venn diagram of unique crashes}
\label{fig:venn-crash}
\end{figure}

\parabf{Crash detection.} We compare the bug finding capabilities of \tech and \titanfuzz{} on an example library \pt using the number of unique crashes as a metric. We did not include inconsistency bugs in this comparison, because crashes are easier to measure and can serve as an approximation of bug finding capabilities~\cite{klees2018evaluating}. We run both tools with their default settings, and execute the programs directly to detect crashes. In addition to direct execution, we also execute all programs with \ad oracles to detect more crashes. Figure~\ref{fig:venn-crash} shows the Venn diagram comparison 
of \tech-\ft/-\zs/-\ft and \titanfuzz where the number in the parenthesis is the number of total crashes found by each technique. First, we observe that all three of our techniques can find more crashes in total (including both \ad and direct execution crashes) compared with the baseline of \titanfuzz, e.g., our default \techfs detects \textbf{2.5} times as many unique crashes as \titanfuzz{}. Moreover, by combining \ft/\zs/\ft together, \tech can detect 19 distinct crashes in total, with 14 unique crashes that cannot be found by \titanfuzz{}, while only 1 crash is uniquely found by \titanfuzz{}, highlighting the effectiveness of \tech{} in generating unusual crash-triggering programs with the help of historical bug-triggering code snippets.

\parabf{Generation efficiency.}
We further discuss the generation efficiency for \tech{} and the strongest baseline \titanfuzz{}. 
Since \titanfuzz{} uses \codex to first generate 25 seed programs and then performs mutations with \incoder, we apply \tech-\fs to only generate 25 programs using \codex (denoted as \tech-\fs-25). According to Table ~\ref{tab:comp-prior-cov}, even \tech-\fs-25 can substantially outperform \titanfuzz{} with much lower cost (\titanfuzz{} further involves additional mutations with \incoder for each API), demonstrating that \llm{s} like \codex can effectively leverage historical bug-triggering programs to generate valuable programs for fuzzing. Moreover, the mutation phase of \titanfuzz{} does not bring significant coverage gain compared to its seed-only version (i.e., \titanfuzzseed), while \tech{'s} coverage does not saturate even after 100 generations (Figure~\ref{fig:cov-trend-full}). Please note that overall it is hard to precisely compare the efficiency of techniques using different \llm{s} (due to different CPU/GPU/Cloud costs), and we tried our best to make the discussion fair here.

\subsection{Ablation Study}
\label{sec:ablation}

\subsubsection{\Fewshot Learning.}
\label{sec:ablationfewshot}

\begin{figure}[t]
\begin{subfigure}[b]{0.49\columnwidth}
    \centering
    \includegraphics[width=0.95\textwidth]{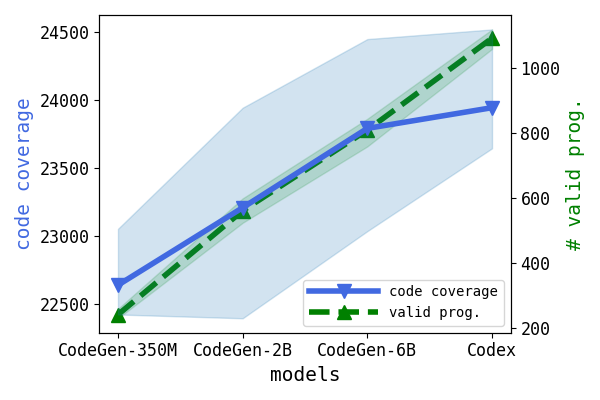}
    \caption{\fewshot}
    \label{fig:fewshot-size}
\end{subfigure}
\begin{subfigure}[b]{0.49\columnwidth}
    \centering
    \includegraphics[width=0.95\textwidth]{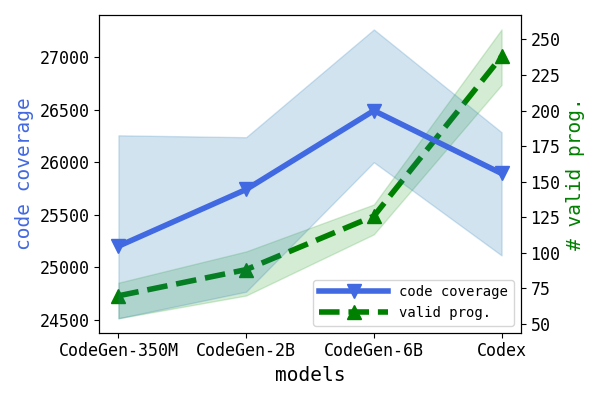}
    \caption{\zeroshot}
    \label{fig:zeroshot-size}
\end{subfigure}
\caption{Comparison of model size scaling}
\end{figure}

We first study the various design choices for \tech-\fs which provide the \llm{s} with several real bug-triggering code examples in the prompt. We compare different models and variants of the prompting strategies.

\parabf{Model size.} We first evaluate the performance of \techfs{} with the model size scaling. Figure~\ref{fig:fewshot-size} plots the code coverage and \# of valid programs generated as we increase model size. We can see that larger models are able to generate more (unique) syntactically and semantically correct programs. We can also observe a clear gain in coverage with the increase of model parameters (from \codegen 350M to 2B and finally 6B), and that \codegen{-6B} can already achieve comparable performance compared to \codex in terms of coverage.

\parabf{Chain-of-Thought prompting.} We now examine the effectiveness of the bug description in our \fewshot template. Our default prompting strategy \textbf{w/ CoT} provides natural language explanation to the buggy code (shown in Figure~\ref{fig:prompt}), which can be seen as chain-of-thought prompting as it instructs the \llm to first generate the possible bug reason (in natural language) and then generate code conditioned on it. The baseline strategy \textbf{w/o CoT} removes the \CodeIn{Bug description: ...} component from the prompt and only asks the model to generate programs from the specified API. As shown in Table ~\ref{tab:torch-fewshot-desc}, including the bug description significantly improves the code coverage, indicating that it is beneficial to give \llm{s} some intermediate context information to ``reason'' about the buggy patterns. Table ~\ref{tab:torch-fewshot-desc} also shows that including natural language description in each example encourages the model to generate more unique APIs, potentially suggesting that the model first generates more diverse natural language description which affects the later code generation. The lower valid rate is probably because the generated programs are more likely to cover some edge-cases and trigger run-time exceptions.

\begin{table}[!htp]\centering
\caption{\tech-\fs w/ or w/o \coft prompting.}\label{tab:torch-fewshot-desc}
\scriptsize
\begin{tabular}{l|rr|rr|r|rr}\toprule
\textbf{Prompt} &\textbf{Valid APIs} &\textbf{All APIs} &\textbf{Valid Prog.} &\textbf{All Prog.} &\textbf{Valid(\%)} &\textbf{Cov} \\\midrule
w/ CoT &\textbf{190} &\textbf{428} &1092 &\textbf{4885} &22.36\% &\textbf{23945} \\
w/o CoT &181 &377 &\textbf{1346} &4798 &\textbf{28.05\%} &22922 \\
\bottomrule
\end{tabular}
\end{table}

\begin{figure}[h]
    \captionsetup{justification=centering}
    \centering
    \includegraphics[width=0.85\linewidth]{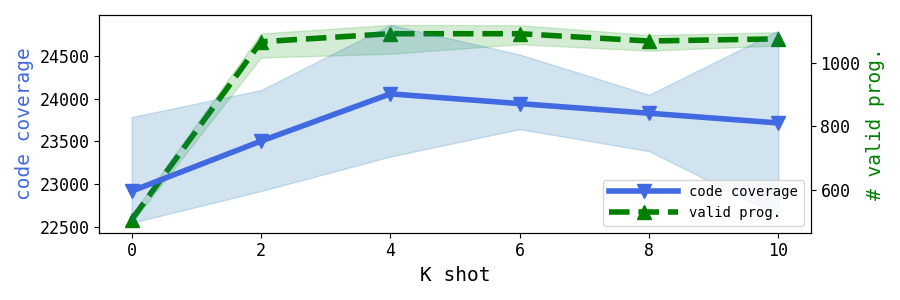}
    \caption{\techfs as the number of examples K increases.}
    \label{fig:kshot-trend}
\end{figure}

 \parabf{\# of examples.} We study the effect of $\kshot$, the number of examples in the context. Figure~\ref{fig:kshot-trend} shows trend of the code coverage and \# of valid programs as we increase the number of examples. We first notice that having no examples ($\kshot=0$) is by far the worst in terms of both coverage and valid programs generated. As we slowly increase $\kshot$, the coverage improves drastically, demonstrating the benefit of using \fewshot learning to provide in-context examples for the \llm to learn from. However, we see that the coverage actually begins to decrease as we add more examples. This could be due to having too many prior historical bug-triggering examples, causing the \llm{s} to restrict its generation creativity and get distracted. In fact, it has been also observed in prior work~\cite{xie2021explanation} that the distracting prompt structure, with more few-shot examples, can decrease the \llm{'s} performance.

\subsubsection{\Zeroshot Learning}
\parabf{Model size.} 
Figure~\ref{fig:zeroshot-size} shows the code coverage and \# of valid programs generated using \techzs as we vary the size of the model used. We first observe a clear trend of improvement as we increase model size, except that the coverage drops for \codex. One reason could be \codex generates much more valid programs, and may miss coverage obtained during exception-handling code (covered by invalid programs).
Additionally, this could also be due to the usage of partial programs in the \zeroshot setting where we directly re-use part of the historical bug-triggering code. As such, smaller models like \codegen can obtain a high coverage results without needing to generate a complete code snippet required in the \fewshot setting.
Nevertheless, we still observe that larger models like \codex is able to achieve the highest number of generated valid programs which are important to test DL libraries.

\newcommand{\editing}{editing\xspace}
\newcommand{\completedft}{completion\xspace}
\newcommand{\completenocode}{completion-NL\xspace}

\parabf{Prompting Strategy.} We evaluate three different \zeroshot prompting strategies: \textbf{editing} (to edit a complete program), \textbf{\completedft} (to complete a partial program, our default \zeroshot variant), and a baseline \textbf{\completenocode} where only a natural language description and no code is given to the \llm for completion. As shown in Table~\ref{tab:torch-zeroshot-prompt}, \completedft achieves significantly higher coverage than \completenocode, highlighting the effectiveness of including partial historical bug-triggering programs in the prompt. It is also noteworthy that \completedft has lower valid rate than 
\completenocode, because the partial code contains unusual patterns and thus is harder to generate semantically valid completions. The extremely low valid rate of \editing is because editing an existing program to use new APIs fully automatically is an even more challenging task for \codex.

\begin{table}[!htp]\centering
\caption{\tech{-\zs} with different prompting strategies.}\label{tab:torch-zeroshot-prompt}
\scriptsize
\begin{tabular}{l|rr|rr|r|rr}\toprule
\textbf{Prompt} &\textbf{Valid APIs} &\textbf{All APIs} &\textbf{Valid Prog.} &\textbf{All Prog.} &\textbf{Valid(\%)} &\textbf{Cov} \\\midrule
\editing &22 &304 &20 &1609 &1.22\% &19440 \\
\completenocode&\textbf{192} &\textbf{400} &\textbf{506} &\textbf{4972} &\textbf{10.17\%} &22917 \\
\completedft &112 &362 &238 &3470 &6.86\% &\textbf{25893} \\
\bottomrule
\end{tabular}
\end{table}

\subsubsection{Fine-tuning.}
\parabf{Model size.} We study the fine-tuning performance with different model sizes. Table ~\ref{tab:comp-torch-finetune-llmsize} shows that \codegen{-6B}, the largest \codegen{} model studied, does achieve the highest code coverage. However, with fine-tuning, larger models are not always better. For example, the 2B model has the most unique valid programs. This could potentially be explained with the validity-unusualness trade-off: as we fine-tune \llm{s} towards unusualness-favored generation, we may lose some validity-preserving information, while both contribute to code coverage and the ultimate bug-finding capability.

\begin{table}[!htp]\centering
\caption{\tech-\ft with different \llm size.}\label{tab:comp-torch-finetune-llmsize}
\scriptsize

\begin{tabular}{l|rr|rr|r|rr}\toprule
\textbf{Model} &\textbf{Valid APIs} &\textbf{All APIs} &\textbf{Valid Prog.} &\textbf{All Prog.} &\textbf{Valid(\%)} &\textbf{Cov} \\\midrule
350M &118 &\textbf{329} &682 &\textbf{4236} &16.10\% &23352.2 \\
2B &118 &292 &\textbf{835} &3393 &\textbf{24.60\%} &22688.8 \\
6B &\textbf{131} &311 &768 &3532 &21.75\% &\textbf{24420.8} \\
\bottomrule
\end{tabular}
\end{table}

\parabf{Prompting strategy.} According to Table~\ref{tab:comp-torch-finetune-prompt}, similar to the \fewshot{} results, including bug description in the fine-tuning process also helps the fine-tuned \llm to generate much more diverse outputs (i.e., more unique and valid APIs/programs) and achieve higher coverage.

\begin{table}[!htp]\centering
\caption{\tech-\ft with different prompting strategies.}\label{tab:comp-torch-finetune-prompt}
\scriptsize
\begin{tabular}{l|rr|rr|r|rr}\toprule
\textbf{Prompt} &\textbf{Valid APIs} &\textbf{All APIs} &\textbf{Valid Prog.} &\textbf{All Prog.} &\textbf{Valid(\%)} &\textbf{Cov} \\\midrule
w/ CoT &\textbf{131} &\textbf{311} &\textbf{768} &\textbf{3532} &\textbf{21.75\%} &\textbf{24420.8} \\
w/o CoT &83 &250 &434 &2659 &16.34\% &24242.8 \\
\bottomrule
\end{tabular}
\end{table}

\subsection{Bug Finding}
\label{sec:bug}
Due to the extensive human cost in bug finding/reporting, in this RQ, we mainly focus on our default setting: \techfs with all the oracles in Section ~\ref{sec:oracle}. Meanwhile, we expect \techzs/\techft to be also effective in bug finding (given their performance in code coverage and crash detection) and may contribute additional bugs.

Bug statistics are summarized in Table~\ref{tab:bugs}. In total, \tech detected \numTotalBugs bugs, with \numTotalConfirmedBugs confirmed, including \numTotalConfirmUnknownBugs confirmed as previously unknown bugs (\numTotalFixedBugs of them have already been fixed).
Besides, Column \textbf{Pending} presents the bugs not yet confirmed and \textbf{Won't Fix} shows bugs rejected by developers (usually due to precision issues or efficiency concerns). Notably, Column \textbf{High Prio} presents the number of high-priority bugs or security vulnerabilities newly detected by \tech. Note that fewer bugs are confirmed or fixed on \tf, because \tf developers are less active (as discussed in Section~\ref{sec:implementation} there are fewer PRs in \tf). Out of the \numTotalConfirmUnknownBugs confirmed new bugs (including \numTotalCrashBugs crashes and \numTotalConsistencyBugs inconsistencies; within them  \numTotalADRelatedBugs are \ad{-related}), only \numTitanFuzzCanFindConfirmedUnknownBugs can be found by running \titanfuzz{} (augmented with our oracles), and \numHistoryBugCanFindConfirmedUnknownBugs can be found by directly rerunning historical bug-triggering programs with our oracles. We next present two exemplary bugs detected by \tech. 

\definecolor{Gray}{gray}{0.85}
\begin{table}[!htp]\centering
\caption{Summary of detected bugs.}\label{tab:bugs}
\scriptsize
\begin{tabular}{lccccc|c}\toprule
&\multirow{2}{*}{\textbf{Total}} &\multicolumn{2}{c}{\textbf{Confirmed (Fixed)}} &\multirow{2}{*}{\textbf{Pending}} &\multirow{2}{*}{\textbf{Won't Fix}} & \multirow{2}{*}{\textbf{High Prio}} \\\cmidrule{3-4}
& &\textbf{Unknown} &\textbf{Known} & & &  \\\midrule
\pt &43 &33 (6) &5 (1) &1 &4 &\numPtHighPrioBugs \\
\tf &33 &16 (0) &7 (0) &5 &5 &\numTfVulnerability \\ \midrule
Total &\numTotalBugs &\numTotalConfirmUnknownBugs (6) &12 (1) &6 &9 &\numTotalHighPrioBugs \\
\bottomrule
\end{tabular}
\end{table}

\begin{figure}
    \centering
    \includegraphics[width=0.9\linewidth]{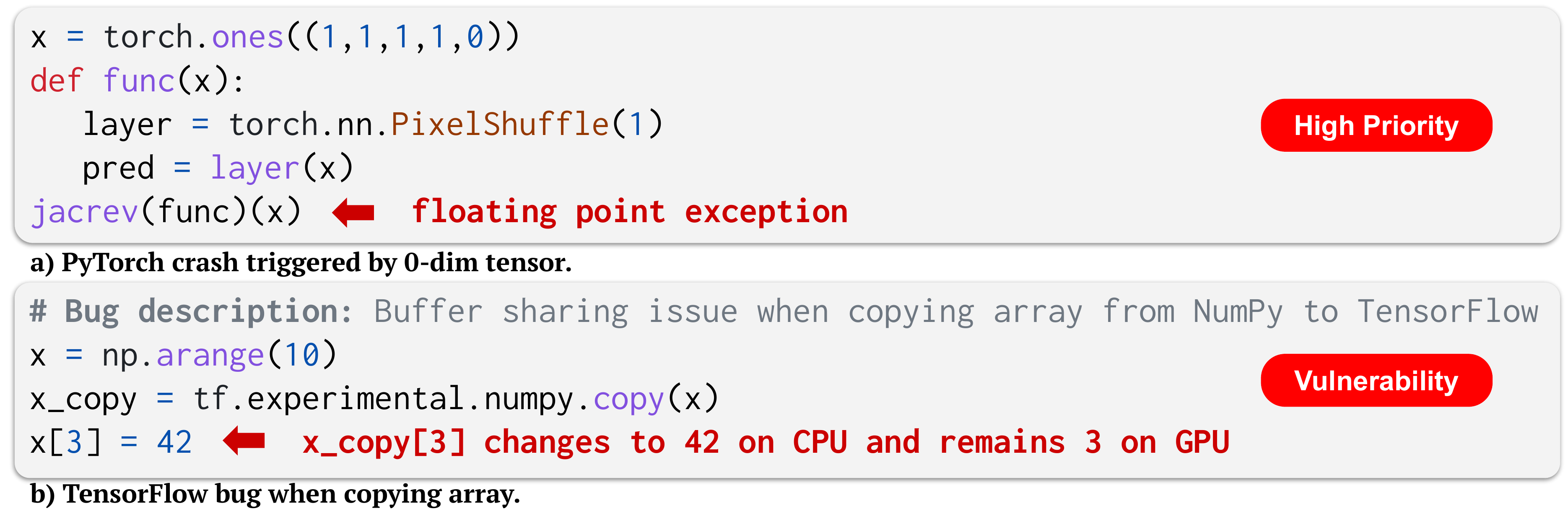}
    \caption{Example bugs found by \tech.}
    \label{fig:examplebugs}
\end{figure}

Figure~\ref{fig:examplebugs}a shows a crash bug when we apply \CodeIn{PixelShuffle} on a special tensor of 0-dimension shape and then compute gradient with \CodeIn{jacrev}. Normally \CodeIn{PixelShuffle} accepts a 4-D tensor where the last dimension is the width of of an image, typically larger than 1. \tech has learned from an in-context historical bug example~\cite{zerodimhistoricalbug} where zero-dimension tensors can trigger crashes, and generates this unusual input of shape \CodeIn{(1,1,1,1,0)} for \CodeIn{PixelShuffle} which triggers floating point exception during gradient computation. As \CodeIn{PixelShuffle} is a commonly-used API in computer vision applications~\cite{shi2016real} where crashes can lead to security risks, this bug is labeled by \pt developers as \textbf{high-priority} and immediately fixed.

Figure~\ref{fig:examplebugs}b presents a \tf bug where the generated fuzzing code snippet first makes a copy of a numpy array \CodeIn{x} and then modifies its value. In this case, the tensor \CodeIn{x\_copy} should remain the same. However, we find that on CPU, after we assign a new value to \CodeIn{x[3]}, the value of copied tensor \CodeIn{x\_copy} is also modified! Interestingly, the bug description (on the first line) generated by \tech seems to correctly capture the reason for this bug! Previous work (including \titanfuzz which also uses \llm{s} for generation) cannot trigger this bug because it requires calling the \CodeIn{copy} API and then modifying the value of the original data - a series of unnatural operations. \tech successfully finds this bug because our \coft prompting instructs it to first predict a plausible bug reason ``buffer sharing issue'' together with its triggering condition ``when copying array ...'' as a bug description. Following this description, \tech can generate the unusual program which is very rare in the training dataset and thus hard to be generated by \titanfuzz. This bug enables data manipulation attacks by silently changing copies of the original array, and has been further confirmed as a \textbf{security vulnerability} by the Google security team.
\section{Discussion}
\label{sec:discuss}

\begin{table}[!htp]\centering
\caption{\tech{ w/o historical information} with \chatgpt.}\label{tab:chatgpt}
\scriptsize
\scalebox{0.85}{
\begin{tabular}{l|rr|rr|r|rr}\toprule
\textbf{System Message:} You are a pytorch fuzzer. &\multicolumn{2}{c|}{\textbf{\# APIs}} &\multicolumn{2}{c|}{\textbf{\# Prog.}} & & \\\cmidrule{2-5}
\textbf{Prompt:} Please generate a program to use conv2d ... &\textbf{Valid} &\textbf{All} &\textbf{Valid} &\textbf{All} &\textbf{Valid(\%)} &\textbf{Cov} \\\midrule
to demonstrate the example usage \textbf{(baseline)} &102 &183 &2298 &4209 &\textbf{55.00\%} &20454 \\ \midrule
in a way you have not seen in your training dataset &190 &297 &\textbf{2306} &4756 &48.00\% &\textbf{21077} \\
in a very creative way &201 &318 &2104 &4801 &44.00\% &20971 \\
in a non-conventional way &\textbf{212} &\textbf{330} &2038 &\textbf{4880} &42.00\% &20759 \\
in a way that is rarely used by developers in practice &200 &306 &2056 &4841 &42.00\% &20759 \\
in a very strange way &181 &291 &1735 &4854 &36.00\% &20377 \\
\bottomrule
\end{tabular}}
\end{table}

\parabf{\tech w/o historical information.} So far we have leveraged historical bug-triggering programs to guide \llm{s} for \unusual program generation. Meanwhile, with the recent advances in the instruct-following capability of \llm{s}, it is also possible to directly instruct \llm{s} (without any historical information) to generate \unusual programs for fuzzing. To this end, we have tested state-of-the-art \chatgpt~\cite{chatgpt} with a list of representative prompts on the 50 \pt APIs used in our ablation study. More specifically, we first instruct \chatgpt to generate typical example usages of an API as the baseline; then, we instruct \chatgpt to generate unusual programs using various other prompts. The shortened prompts and main results have been shown in Table~\ref{tab:chatgpt}. From the table, we can observe that all the studied prompts can help \chatgpt cover much more APIs than the baseline, demonstrating \chatgpt and similar models (e.g., GPT-4~\cite{openai2023gpt4}) can understand the instructs and generate more interesting programs that may cover interesting library paths/behaviors. Another interesting observation is that the other prompts all have lower valid rates than the baseline, since less common programs may more likely fail.

\parabf{Impact of example selection.} 
Besides the random example selection used in our default \techfs variant, we have also investigated other strategies to select in-context examples for few-shot fuzzing. Intuitively, examples with APIs semantically/syntactically similar to the target API may provide more relevant bug-triggering patterns that can be better utilized by \llm{s}. Conversely, a diverse set of examples can also provide complimentary bug-triggering patterns/ingredients that could be leveraged/combined by \llm{s}. As such, we design a set of smoothed maximum-marginal-relevance (MMR)~\cite{mmr} guided selection strategies inspired by prior work ~\cite{ye2022complementary}, including strategies favoring APIs similar to the target API, strategies favoring diverse APIs, and strategies in-between. Interestingly, we observe that the default random strategy is competitive compared with all studied variants. The main reason could be that modern \llm{s} are powerful enough to learn even from dissimilar examples for program generation; in this way, random selection can provide a diverse set of ingredients to facilitate effective generation.   

\parabf{Threats to validity.} The main threats to internal validity lie in the potential bugs in our implementation and experimentation. To mitigate such threats, {we} performed rigorous review for our code. The main threats to external validity lie in the subject systems used. To reduce the threats, we select two most popular DL libraries, \pt and \tf, which have also been widely studied in recent work~\cite{freefuzz,deeprel,nablafuzz,titanfuzz,audee}. Lastly, we also adopt widely used metrics in prior fuzzing work~\cite{freefuzz,nablafuzz,titanfuzz}, such as real bug detection and code coverage.

\section{Conclusion}
We have introduced \tech, the first approach to leveraging historical bug-triggering programs to prime \llm{s} for fuzzing with edge cases. Compared to traditional fuzzing techniques on leveraging such historical information studied for over a decade, \tech is fully automated, generalizable, and applicable to challenging domains, such as DL library fuzzing. Moreover, \tech also shows the potential of \chatgpt for edge-case program generation without any historical information. The experimental results show that \tech substantially outperforms existing DL library fuzzers, and can detect various bugs for \pt and \tf. 

\parabf{Artifact Availability.} We make our artifact available at ~\cite{fuzzgptcode}.

\bibliographystyle{ACM-Reference-Format}
\bibliography{references}

\end{document}